\documentclass[12pt]{article}

\def\hybrid{\topmargin -20pt  \oddsidemargin 0pt
      \headheight 0pt   \headsep 0pt
      \textwidth 6.25in % A4 paper
      \textheight 9.5in % A4 paper
      \marginparwidth .875in
      \parskip 5pt plus 1pt   \jot = 1.5ex}

\hybrid

\usepackage{amssymb}

\def\beqa{\begin{eqnarray}}
\def\eeqa{\end{eqnarray}}

\def\x{\times}
\def\ox{\otimes}
\def\o+{\oplus}
\def\ra{\rightarrow}
\def\lra{\longrightarrow}

\def\Lra{\Longrightarrow}

\def\Llra{\Longleftrightarrow}
\def\hra{\hookrightarrow}
\def\da{\downarrow}

\newcommand{\ov}{\overline}
\newcommand{\un}{\underline}
\newcommand{\del}{\partial}
\def\delbar{\bar{\partial}}

\newcommand{\al}{\alpha}
\newcommand{\be}{\beta}
\newcommand{\la}{\lambda}
\newcommand{\si}{\sigma}
\newcommand{\om}{\omega}

\newcommand{\A}{{\cal A}}
\newcommand{\C}{{\cal C}}
\newcommand{\D}{{\cal D}}
\newcommand{\E}{{\cal E}}
\newcommand{\cF}{{\cal F}}
\newcommand{\G}{{\cal G}}
\newcommand{\cH}{{\cal H}}
\newcommand{\J}{{\cal J}}
\newcommand{\cL}{{\cal L}}
\newcommand{\M}{{\cal M}}

\newcommand{\cO}{{\cal O}}
\newcommand{\cP}{{\cal P}}
\newcommand{\R}{{\cal R}}
\newcommand{\cS}{{\cal S}}
\newcommand{\T}{{\cal T}}
\newcommand{\cW}{{\cal W}}

\def\Pfaff{\mbox{Pfaff}}

\newcommand{\resetcounter}{\setcounter{equation}{0}}

\begin{document}
\thispagestyle{empty}
\rightline{LMU-ASC 17/09}
\vspace{2truecm}
\centerline{\bf \LARGE Perspectives on Pfaffians}
\vspace{.5truecm}
\centerline{\bf \LARGE of Heterotic World-sheet Instantons}

\vspace{1.5truecm}
\centerline{Gottfried Curio\footnote{gottfried.curio@physik.uni-muenchen.de}} 

\vspace{.6truecm}

\centerline{{\em Arnold-Sommerfeld-Center for Theoretical Physics}}
\centerline{{\em Department f\"ur Physik, 
Ludwig-Maximilians-Universit\"at M\"unchen}}
\centerline{{\em Theresienstr. 37, 80333 M\"unchen, Germany}}

\vspace{1.0truecm}

\begin{abstract}
To fix the bundle moduli of a heterotic compactification
one has to understand the Pfaffian one-loop prefactor
of the classical instanton contribution.
For compactifications on elliptically fibered Calabi-Yau spaces $X$
this can be made explicit for spectral bundles 
and world-sheet instantons supported on rational base curves $b$:
one can express the Pfaffian in a closed algebraic form as a polynomial, 
or it may be understood as a $\theta$-function expression. 
We elucidate the connection between these two points of view
via the respective perception of the relevant spectral curve,
related to its extrinsic geometry in the ambient space 
(the elliptic surface in $X$ over $b$) or 
to its intrinsic geometry as abstract Riemann surface.
We identify, within a conceptual description, general vanishing loci
of the Pfaffian, and derive bounds on the vanishing order, relevant
to solutions of the equations $W=dW=0$ for the superpotential.

\end{abstract}

\newpage

\section{Introduction}

In a heterotic string compactification on a Calabi-Yau space $X$ with bundle
$V$, giving $N=1$ supersymmetry in the effective 4D theory,
world-sheet instantons can generate an effective superpotential
for the moduli, thus potentially
partially lifting the classical moduli space.\footnote{we 
treat a single instanton contribution; the
sum might vanish non-trivially in some situations}
This involves a moduli-dependent one-loop functional determinant.
We will study world-sheet instantons,
wrapping (once) smooth holomorphic curves $b$ in $X$.
As contributions to the superpotential $W$ arise only at tree-level
in string perturbation theory, a contributing $b$ has genus zero. 
As just two fermion zero-modes on $b$ are needed to generate 
a superpotential, one has, conjecturally\footnote{it is not clear 
that additional fermion zero-modes could not be lifted by higher-order
interactions}, as further necessary condition that $b$ is isolated 
(otherwise additional fermion zero-modes on $P$ arise)\footnote{Instead 
of generating a superpotential world-sheet instantons
may also deform the complex structure of the classical moduli space;
this possibility is realized if $b$ moves in a family [\ref{Wfam}].}.
So we assume throughout that $b$ is a smooth, isolated rational curve.
Actually we take $X$ elliptically fibered $\pi:X\ra B$ 
(with section $\sigma$ and fibre $F$) and $b$ a curve in the base $B$ 
(embedded via $\sigma$).
For $SU(n)$ bundles $V$ which arise (for all details on the notions and 
notations of the spectral cover method cf.~app.~\ref{data})
as $V=p_*(p_C^*L\otimes \cP)$ from a spectral cover surface $C$
(an $n$-fold ramified cover of $B$) endowed with a line bundle $L$ 
(say, arising via restriction from a line bundle $\underline{L}$ on $X$)
one has a cover curve $c$ of $b$ and 
a line bundle $l=\underline{L}|_{\E}$ on $\E=\pi^{-1}(b)$ with 
$V|_b\cong \pi_{c*}l|_c$.
Let us denote $l(-F)=l\ox \cO_{\E}(-F)$ by $\widetilde{\cL}$, 
and let $\cL:=\widetilde{\cL}|_c$
and $\cL_{c'}:=\widetilde{\cL}|_{c'}$ for $c'\in |c|$
(the system of linear equivalent divisors)
or $\cL_{c_t}$ for $t \in \M_{\E}(c)$.

To understand on a computational level 
the moduli dependence of the Pfaffian prefactor 
$Pfaff$ in the superpotential $W_b$ 
caused by a world-sheet instanton supported on $b$
(cf.~sect.~\ref{section of line bundle})
we note that, from different perspectives, 
two equivalent expressions arise
\begin{itemize}
\item a polynomial expression in algebraic moduli 
(parameters describing motion in $|c|$)
\item a $\theta$-function expression in transcendental moduli
\end{itemize}
The two ensuing expressions for $Pfaff$ are of course related with each other.

Recall that a non-trivial contribution $W_b$ of $b$ 
to the world-sheet instanton superpotential occurs precisely for
$H^0(b, V|_b\ox \cO_b(-1))=0$. So
the vanishing locus of $Pfaff$ is given by the vanishing locus of 
an expression which controlles the non-triviality of $H^0(c, \cL_{c_t})$
as $t$ varies over the relevant moduli space.
(This leads, via arguments of holomorphy and a consideration of the possible
power occurring, to an identification of the Pfaffian with such a
'controlling expression', typically a certain determinant, cf. below).

Now let us come back to the expressions for the Pfaffian.
The first expression, giving the polynomial $\det \iota_1$, 
arises via the {\em extrinsic} algebraic 
exact\footnote{we made the technical assumption $\la > 1/2$ 
on the spectral twist parameter $\la$
and assume the necessary condition 
$h^0(\E, \widetilde{\cL}\otimes \cO_{\E}(-c))=h^0(\E, \widetilde{\cL})$
for contribution of $b$ to $W$ to be fulfilled (sect.~4.2 [\ref{extrinsic}])} 
sequence
\beqa
\label{extrinsic sequence}
0\lra H^0(c, \cL)\buildrel\delta \over \lra 
H^1(\E, \widetilde{\cL}\otimes \cO_{\E}(-c))
\buildrel\iota_1 \over \lra H^1(\E, \widetilde{\cL})\lra H^1(c, \cL)\lra 0 \;\;
\eeqa
with $\iota_1$ induced from multiplication with a moduli-dependent element 
$\tilde{\iota}\in H^0(\E, \cO_{\E}(c))$. On the other hand 
one has also the {\em intrinsic} transcendental exact sequence
\beqa
\label{intrinsic sequence}
0\lra H^0( c, \cL) \lra \C^{\infty}(c, \cL)\buildrel \bar{\partial} \over \lra
\C^{\infty}(c, \cL \otimes \Omega^{0,1}_c)\lra H^1( c, \cL) \lra 0\;\;
\eeqa
where the determinant of the $\bar{\partial}$ operator gives rise to 
the $\theta$-function expression. We connect these two points of view
by showing how the $\theta$-function expression arises also in the
algebraic framework, cf.~(\ref{theta Pfaff relation}). 
Further the sequences are interwoven with each other
\beqa
\begin{array}{ccccccccc}
&&0&&0&& 0 && \\
&&\da&&\da&& \da &&\\
0&\ra&H^0(\E, \widetilde{\cL}\otimes \cO_{\E}(-c)) &\ra & 
H^0(\E, \widetilde{\cL})&\ra&H^0(c, \cL)&\buildrel\delta\over\ra & \\
 & & \da & & \da \;\; & & \da & & \\
0 & \ra & \C^{\infty}(\E, \widetilde{\cL}\otimes \cO_{\E}(-c))& \ra & 
\C^{\infty}(\E, \widetilde{\cL}) & \ra & 
\C^{\infty}(c, \cL)&  & \\
 & & \;\;\;\; \da \bar{\partial}
& & \;\; \da \bar{\partial}& & \;\;\; \da \bar{\partial} & & \\
0 &\ra & \C^{\infty}(\E, \widetilde{\cL}\otimes \cO_{\E}(-c)
\otimes \Omega^{0,1}_{\E})
&\ra & \C^{\infty}(\E, \widetilde{\cL}\otimes \Omega^{0,1}_{\E})
&\ra & \C^{\infty}(c, \cL\otimes \Omega^{0,1}_c)
&  & \\
& & \da & & \da & & \da & & \\
& \buildrel \delta \over \ra & H^1(\E, \widetilde{\cL}\otimes \cO_{\E}(-c)) 
&\buildrel\iota_1\over\ra &H^1(\E, \widetilde{\cL})&\ra&H^1(c, \cL) & \ra & 0\\
 & &  & &  & & \da & & \end{array}\nonumber\\
 \hspace{4.7cm}  \hspace{3.7cm} 0 \hspace{2.75cm} 
\eeqa
Here the first two $H^0$-groups vanish if $b$ is able to contribute to the
superpotential at all, cf.~(\ref{neces cond}), a necessary condition
we assume to be fulfilled throughout.

The algebraic sequence arises {\em horizontally}, the transcendental
one {\em vertically}; when considering the moduli space fibration 
there will be a second reason for these denotations.

A $\theta$-function representation for $Pfaff$ arises 
generally from the transcendental sequence (\ref{intrinsic sequence}),
cf.~sect.~\ref{theta-function representation}.
After recalling how the algebraic expression arises from
(\ref{extrinsic sequence}) we show how a $\theta$-function arises also
directly in the algebraic approach, cf.~(\ref{theta Pfaff relation}). 

Because the chiral Dirac operator 
is closely related to the $\bar{\partial}$-operator one can approach
all statements within the algebraic geometric category. In such a framework
the (zero) divisor given by the vanishing locus of $Pfaff$ in the moduli space
will be crucial. We continue our study [\ref{extrinsic}]
of vanishing loci of $Pfaff$. The codimension zero
loci are given by those components of the reducible
moduli space for which the purely topological criterion for non-contribution
(of the world-sheet instanton supported on $b$)
to the superpotential applies. In the next codimension the mentioned
vanishing divisor occurs;
this divisor will in general be neither irreducible nor will its components be
reduced: for corresponding examples of $Pfaff=fg$ or $Pfaff=fg^k$ 
cf.~[\ref{extrinsic}], [\ref{BDO}] 
and sect.~\ref{more specialised information}. 
In these examples we could give explicit factors of $Pfaff$ 
from structural reasons [\ref{extrinsic}]. More generally
the corresponding loci arising from such structural arguments have
higher codimension; they are specific subvarieties of the vanishing
divisor which have an explicit description, cf.~(\ref{Sigma remark 3});
we will give the corresponding nested hierarchy of loci, 
cf.~sect.~\ref{loci} and \ref{coordinates};
it would be interesting to
compare these algebraic geometric descriptions with correspon\-ding facts 
on the side of the explicit transcendental $\theta$-function.

Furthermore the mentioned loci are typically of higher multiplicity.
This is (besides the $\theta$-function)
the other topic considered here which goes beyond the considerations
in [\ref{extrinsic}]. Although we will not touch the question 
of summing up the world-sheet instanton superpotential from
its different contributions (supported on different curves)
we consider the question of solutions to $W=dW=0$ on the 
level of an individual instanton contribution: what is meant here
comes down to studying the multiplicity $k$ (later denoted by $k'$)
with which a factor $f$ occurs
in $W=f^k g$, in particular whether it is $\geq 2$.
We will study the question of multiplicity in some generality before
we apply this to the two main examples of [\ref{BDO}] (which were also treated
in [\ref{extrinsic}]) and the generalisation given in [\ref{extrinsic}], 
cf.~sect.~\ref{multiplicities} and \ref{The SU(3) case},
deriving, in particular, conceptually
that multiplicities $\geq 2$ occur there. This supplements and completes our
conceptual treatment of the factor $f^k$ in [\ref{extrinsic}].

In {\em sect.~\ref{section of line bundle}} 
we recall the definition of the Pfaffian prefactor
and some subtleties connected with it. We also point to the interpretation
of $Pfaff$ as a section of the square root of the determinant line bundle
for a family of complex chiral Dirac operators. This usually occurs for
a family of varying curves (with associated Dirac or $\delbar$ operators);
in our case the support curve $b$ of the world-sheet instanton stays fixed
and the bundle over it varies; for our case of a base curve $b\subset B$
for spectral cover
bundles over an elliptically fibered Calabi-Yau space $X$ this movement in 
bundle moduli space comes down again to the variation of the spectral cover
curve $c$ which lies in the elliptic surface $\E=\pi^{-1}(b)$ over $b$. 
In this connection we also point to the general
theta-function expression arising in such a set-up.
In {\em sect.~\ref{Spectral bundles over X}, \ref{restricted bundles}}
and {\em \ref{Bundles over b}} we study in detail the bundle moduli spaces
over $X$, $\E$ and $b$. We will assume that there are no continuous moduli 
for line bundles over the spectral cover surface $C\subset X$, lying over $B$ 
(in the moduli space fibration they correspond to a fibre direction, 
cf.~[\ref{CD}]).
Then the bundle moduli are given, 
up to a discrete parameter $\la\in \frac{1}{2}{\bf Z}$ (still a fibre direction
in moduli space),
by the movements of $C$ in $X$ (horizontal base direction in moduli space). 
By restriction corresponding statements ensue 
for the case of $c\subset\E$, lying over $b$. Here, however, the 
moduli space for spectral bundles which are defined a priori over $\E$
(not arising by restriction from $X$) has 
also continuous fibre direction moduli as here the Jacobian of $c$ enters the
story. In sect.~\ref{loci} - \ref{The SU(3) case} 
we give the nested hierarchy of vanishing loci, cf.~(\ref{Sigma remark 3}), 
discuss the question of multiplicity
and put examples into this context in sect.~\ref{more specialised information}.
In {\em sect.~\ref{intrinsic}} we demonstrate how the theta-function arises
directly in the algebraic geometric approach and compare
the moduli space description of the restricted bundle $V|_{\E}$ and
the corresponding universal Jacobian fibration over the moduli space $\M_g$
of curves where the general transcendental theta-function lives naturally;
we conclude with remarks on multiplicities. 
The appendices collect various technical material related to the spectral
cover construction (the algebraic approach, app.~A) and 
the Riemann theta function (the transcendental approach, app.~B).

\subsection{Overview and summary}

As the issue in question - the vanishing behaviour of the Pfaffian prefactor
(of a world-sheet instanton superpotential) in dependence 
on the vector bundle moduli - 
necessarily uses a heavy amount of algebraic-geometric notions it may be 
useful to provide here also a nontechnical overview of the more detailled
investigations which follow in the later sections
(herein we allow ourselves to give an only approximate description of various
issues whose potential subtleties are dealt with in the main text).

Clearly one first has to understand the moduli in question.
As the bundle decomposes fibrewise in a sum of line bundles,
each of which can be represented as 
$\cO_F(q_i-p_0)$ by a fibre point $q_i\in F$,
the bundle corresponds to a $n$-fold cover surface $C$ of $B$,
or a corresponding curve $c$ over $b$. 
If $s=0$ is an equation for $c$ in $\E=\pi^{-1}(b)$
then it is the variation of these zerodivisors $(s)$ in the linear
system $|c|={\bf P}H^0(\E, \cO_{\E}(c))$ which gives the variation of the 
bundle $V$ (concretely this comes down to having different polynomial
coefficients $a_i$ in the affine expression $s=a_0+a_2 x+a_3 y + \dots$).

In addition the construction 
can be twisted by a line bundle $\cL=\cO_{\E}(D)$ on $c$ of degree $g_c-1$; 
so one has a representation
\beqa
\cL&=&K_c^{1/2}\ox \cF
\eeqa
where $\cF$ is a flat line bundle on $c$.
Actually there is a universal expression for $\cL$ as coming from a 
restriction to $c$ of a line bundle $\widetilde{\cL}$ on $\E$
(here occur some subtleties concerning the question of
integrality versus half-integrality of cohomology classes
for different parities of parameters involved); one has $\cF=\Lambda|_c^{\la}$
with $\Lambda = \cO_{\E}(ns- (r-n)F)$ and $\la$ a half-integer.

The main contribution criterion states that $Pfaff$ vanishes just if
$V|_b\ox\cO_b(-1)$ has nontrivial sections 
\beqa
\label{main contribution criterion}
Pfaff(t)=0 & \Llra & \Gamma\Big(b, V|_b\ox\cO_b(-1)\Big)\neq 0
\eeqa
what translates for spectral
bundles to the question whether $\cL$ has nontrivial sections.
So, for example, one finds that on the locus $\Sigma$ in the moduli space
$\M_{\E}(c)$
(given by the different concrete curves $c_t$ in $|c|$)
where $\Lambda|_{c_t}$ becomes trivial the divisor $D$ of $\cL$ becomes
effective and so $Pfaff$ vanishes (various refinements of this locus $\Sigma$
will be investigated)
\beqa
\Sigma & \subset & (Pfaff)
\eeqa

The description of this locus is on the one hand satisfying
as it is defined purely structurally. On the other hand one would
also like to connect this with explicit descriptions. 
The latter means an explicit expression given directly in the 
moduli (the coefficients of the $a_i$).
This is accomplished by the definition of the locus $\R$
where all the resultants $R_i^{(j)}=Res(a_i, a_n^{(j)})$ vanish
(with $i=2, \dots, n-1$ and $a_n=\prod_j a_n^{(j)}$)
which will be shown to be a sublocus of $\Sigma$
\beqa
\R & \subset &\Sigma
\eeqa

For our main case of interest, the case of $SU(3)$ bundles (where also
all the main examples occur), one gets even equality between these
subloci in moduli space, giving a precise {\em explicit} description 
of the locus $\Sigma$ in the vanishing divisor $(Pfaff)$ of $Pfaff$
\beqa
\R & = & \Sigma\;\;\;\;\;\;\;\;\;\;\;\;\;\;\;\;\;\; \mbox{for $SU(3)$ bundles}
\eeqa

We can be even more concrete: for $SU(3)$ bundles (as for all $SU(n)$ bundles
with $n$ odd) the half-integral parameter $\la$ in the spectral construction
has actually to be strictly half-integral. As we have furthermore the
standing technical assumption $\la > 1/2$ (cases with $\la < - 1/2$
can be reduced to this situation) actually the first interesting case is
$\la = 3/2$ which is also a major case with regard to the concrete applications
we have in mind (in view of some concretely known examples). In this case
a whole component $(f)$ of $(Pfaff)$ could be described (i.e. a sublocus of
codimension zero; this comes from a relation $f|Pfaff$) which also contains
the locus $\R=\Sigma$. 

When one combines this with certain insights one can gain structurally 
about the multiplicity $k$ with which $f$ occurs in $Pfaff=f^k g$
one can recover the decisive assertions in most of the main examples 
known of the factorisation phenomenon for $Pfaff$ as will be described in 
sect.~\ref{more specialised information}.

Finally the whole discussion can be described also 
from the internal perspective of the algebraic curve $c$.
Then the different moduli are related to the moduli of the abstract 
Riemann surface $c$ and the flat line bundle $\cF_t$ gives for each
concrete curve $c_t$ a point in the Jacobian $Jac(c_t)$.
As both vanishing divisors in the appropriate moduli spaces are related to 
the effectivity of a certain divisor (or the existence of a nontrivial section)
it turns out that $(Pfaff)$ is most closely related to the theta divisor
$\Theta$. This is nice because it shows how the transcendental 
theta function, over a certain locus, can be expressed - 
if one switches from the transcendental period variables 
suitable to the universal Jacobian fibration over the moduli space 
of curves to the algebraic polynomial parameters described earlier - 
as a finite determinant 
(the connection can be deepened for the multiplicities).

\section{\label{section of line bundle}The Pfaffian prefactor}

\resetcounter

The first level of question is whether a specific
world-sheet instanton does make at all a contribution to the superpotential
or not. In the positive case  one wants of course to have more specific
information about the prefactor $f$ to the exponential instanton contribution
\beqa
\label{superpotential}
W_b&=&f\cdot \, \exp\Big(\, i\int_b \tilde{J}\, \Big)
\eeqa
Here $\tilde{J}=B+iJ$ is the K\"ahler class with
$\mbox{area}(b)=2\pi \al' \int_b J$. The prefactor $f$
stems from the one-loop integral over quantum fluctuations 
around the classical instanton solution giving the 
exponential term.\footnote{We will not give a full elucidation of the 
instanton contribution which is somewhat subtle 
as the different naive factors of the moduli space are not decoupled. 
The pfaffian is a section of a line bundle which can be non-trivial 
if $V$ leads to a non-vanishing fivebrane class $W_5\neq 0$;
however, even for $W_5=0$, there is by the lack of a canonical 
trivialization no well-defined phase of the pfaffian [\ref{Wittenhet}]
(its absolute value can be defined by zeta-function regularization);
the phase of the classical exponential (K\"ahler) factor 
has a non-trivial interference with the bundle moduli [\ref{W99}],
yet the total phase in (\ref{superpotential}) is well-defined; 
another subtlety is that the stability notion for the bundle depends on
the K\"ahler class employed.} We recall here some generalities relevant to $f$.

With the kinetic Lagrangian 
$S_{\psi}=\int_b tr\, \bar{\psi}\, D_-\psi \; d\si d\tau$
the fermionic partition function gives 
$\int \D \psi \,\D \bar{\psi} \; e^{-S_{\psi}}\, \sim \, \det \, \D_F$
($\D_{B/F}$ the kinetic term for bosonic/fermionic fluctuations)
in Euclidean space with Weyl-fermions.
$\D_F$ decomposes on $S\ox V|_b$ according to 
the chirality decomposition of the spinor bundle $S=S_+\oplus S_-$ on $b$:
in the left-moving sector one has the chiral Dirac operator 
$D_-:\, S_-\ox V|_b \, \lra \, S_+\ox V|_b$ (with gauge connection $A$)
and on the other hand $i\del_+:\, S^+\ox V|_b \, \lra \, S^-\ox V|_b$ 
(without $A$).

In the decomposition 
$\det\, \D_F =  |\, \det\, \D_F \, | \; e^{i\phi}$
the absolute value is evaluated (with a universal positive 
proportionality constant) via 
$|\, \det\, \D_F \, |^2 \, = \, \det\; \big(\, \D_F\, \D_F^{\dagger}\big) 
\; \sim \; \det \; \D \; = \; \det \; \big( D_-D_-^{\dagger}\big)$
which is a gauge-invariant quantity (here $D_+=D_-^{\dagger}$ replaces 
the $i\del_+$ above, leading to $\D$ replacing the $\D_F$ above);
this leads\footnote{in 2, or more generally $8k+2$, 
dimensions the determinant line
bundle of the complex chiral Dirac operator admits a canonical
square root and $\sqrt{\det \D}$ is a section} 
to $ |\, \det\, \D_F \, | \sim \sqrt{ \det \D}$.

Converting the treatment of Weyl-fermions in Euclidean space 
to the physical case of Majorana-Weyl fermions in Minkowski signature
halves the number of degrees of freedom and leads to taking the square root
$\sqrt{\det\, \D_F}\; = \; \Pfaff (\D_F)$. Then the prefactor $f$ in 
(\ref{superpotential}) can be evaluated as [\ref{Wittenhet}]
(the prime stands for omission of zero-modes) 
\beqa
\label{f}
f&=&\frac{\Pfaff \, '\, (\D_F)}{\sqrt{\det '\, \D_B}}\;\; = \; \;
\frac{\Pfaff\, (\delbar_{V|_b(-1)})}
{(\det\, \delbar_{\cO(-1)})^2 (\det'\, \delbar_{\cO})^2}
\eeqa

The world-sheet instanton computation is often considered in 
the physical gauge. 
The bosons, describing fluctuations of $b$ in its ten-dimensional
ambient space, are $\cO^2\oplus N$-valued, using a complexification 
of four-dimensional extrinsic space and the normal bundle $N$ of $b$ in $X$.
For a smooth rational curve which is isolated one gets 
$N=\cO(-1)\oplus \cO(-1)$ for the normal bundle.
Interpreting the real eight-dimensional normal bundle
as complex four-dimensional one may write $\sqrt{\det \, \D_B}=\det \, \D'_B$.

To describe the world-volume fermions consider the spinor bundle 
$S=S_{-}(b)\oplus S_{+}(b)=K^{1/2}_b\oplus K^{- 1/2}_b$ on $b$ 
(with its left/right-movers decomposition and
after chosing an appropriate complex structure) 
with respective kinetic operators $\delbar$ and $\del$.
The right-moving fermions are sections of
$S_+(b)\ox S_+(\cO^2\oplus N)$
where $S_+(\cO^2\oplus N)=\Big( S_+(\cO^2)\ox S_+(N)\Big) 
\oplus \Big( S_-(\cO^2)\ox S_-(N)\Big)$ is the positive chirality spin bundle
associated to $N$.

Because of unbroken supersymmetry in the instanton field
there is a full cancellation in the right-moving sector
between fermions contributing to $Pfaff'(\D_F)$
and bosons contributing to $\det'\, \D_B$.
The rewriting in (\ref{f}) in the remaining left-moving sector 
(retaining just $\delbar$ operators)
results from the identification
\beqa
\mbox{left-moving fermions} & \cong & 
\Gamma\Big( b , S_-(b)\otimes V|_b\Big)
\eeqa
$D_-$ becomes $(i)\delbar$ (for a suitable
choice of complex structure) and the left-handed spin bundle $S_-(b)$ becomes 
$\cO_b(-1)$ (we assume structure group $SU(n)\subset SO(2n)$ with $n\leq 8$).

The contribution criterion states that $W_b\neq 0$ 
just if $\delbar|_{V|_b(-1)}$ has a zero kernel $H^0(b, V|_b(-1))$. 
If not $H^0(b, V|_b(-1))=0$ everywhere in moduli space $\M$, 
i.e.~if $Pfaff \not\equiv 0$,
it will give in our spectral case a divisor defined by $\det \, \iota_1=0$
in $\M$, where $Pfaff=0$, such that
$Pfaff=(\det \iota_1)^m$ up to a constant (actually $m=1$ [\ref{BDO}]).

\subsection{\label{theta-function representation}
The Pfaffian as a section of a line bundle}

The Pfaffian is a section of the canonically existing square root of 
the determinant line bundle for a family of complex chiral Dirac operators.
A holomorphic family of complex curves $X_m$ (with $m\in \M$) 
gives a family of bundles
\beqa
\label{family of bundles}
{\cal X}\;\;\;\;\;\;\;\;\;\;\;\;\;\;\;\;\;\;\;\;\;\;\;\nonumber\\
\hspace{3cm}\da \; S_-(X_m)\ox V_{X}\\
\M \;\;\;\;\;\,\;\;\;\;\;\;\;\;\;\;\;\;\;\;\;\;\;\nonumber
\eeqa 
and thereby a family $(\delbar_m)_{m\in \M}$ of $\delbar$-operators 
on $\C^{\infty}(X_m, V)$. From the sequence
\beqa
0 \lra H^0(X_m, V)\lra \C^{\infty}(X_m, V)\stackrel{\delbar_m}{\lra}
\C^{\infty}(X_m, V\ox \Omega_{X_m}^{0,1})\lra H^1(X_m, V)\lra 0
\eeqa
one gets a family of one-dimensional vector spaces 
\beqa
\mbox{DET}(\delbar_m)\; & \; = \; & \; 
\Lambda^{h_0}H^0(X_m, V)\ox \Lambda^{h_1}H^1(X_m, V)^*
\eeqa
These complex lines fit together holomorphically over $\M$
to give the determinant line bundle $\mbox{DET}(\delbar)$ 
(endowed with a suitable norm)
over $\M$ with fibre $\mbox{DET}(\bar{\partial}_m)$.

The varying family of $\delbar$ operators may also be obtained 
if the relevant bundle changes instead of the curve.
As in the spectral cover approach the bundle is encoded (essentially)
again by a spectral cover {\em curve} 
this will lead back in the end again to a 
curve $c$, varying over its moduli space (cf.~the space
${\bf P}H^0(\E, \cO_{\E}(c))$ below in sect.~\ref{restricted bundles}).

{\em Remark:} 
The theta function can be understood as the determinant of an appropriate 
$\delbar$ operator: choosing first a spin structure, 
i.e.~a theta characteristic $K_c^{1/2}$,
allows an identification $\phi: \, Pic_{g-1}(c)\, \lra \, Jac(c)$
via the relations (where $\mbox{Div}_{g-1}^{\, eff}(c)=Sym^{g-1}\, c$) 
\beqa
\mbox{Div}_{g-1}^{\, eff}(c)\, \ra \, 
\Big(\mbox{Div}_{g-1}^{\, eff}(c)/\sim\Big) 
\hra  \Big(\mbox{Div}_{g-1}(c)/\sim \Big) \cong  Pic_{g-1}(c)
\stackrel{\phi(\cdot)=\cdot\otimes K_c^{-1/2}}\lra 
\;\;\;\;\;\;\;\;\;Pic_0(c)\nonumber\\
\;\;\;\;\;\;\da \mu \hspace{11cm}
\da \cong \nonumber\\
W_{g-1} \hspace{6.6cm}
\;\; \stackrel{\widetilde{\phi}(\cdot)=\cdot - \mu(K_c^{1/2})}{\lra} 
\;\Theta_c \hra Jac(c)
\nonumber
\eeqa
mapping a degree $g-1$ line bundle $\cL=\cO_c(D_{g-1})=K_c^{1/2}\ox\cF$ 
to the flat bundle $\phi(\cL)=\cF$
(for all the standard notation 
cf.~app.~\ref{Mathematical Standard Notation}).
For the family over $Jac(c)$ of $\delbar$ operators 
\beqa
\C^{\infty}\Big(c, \cO_c(D_{g-1})\Big) & \stackrel{\delbar}{\lra} &
\C^{\infty}\Big(c, \cO_c(D_{g-1})\ox \Omega^{0,1}\Big)
\eeqa
one has, noting that $\theta(\cdot, \Omega_c)$ vanishes just if
$\ker \, \delbar\; = \; H^0(c, \cO_c(D_{g-1}))$ is non-trivial, 
\beqa
DET\;\, \delbar_c \; &  \; = \; & \; \cO_{Jac(c)}\,\big(\Theta_c\big)
\eeqa
Note that this concerns the 'vertical' direction $z$ related to the 
line bundle or divisor and not the 'horizontal' direction related to the 
parameter $\Omega_c$ of the varying curve, cf.~above.

\section{\label{Spectral bundles over X}
The case of base curves in elliptically fibered $X$ 
and fibrewise semistable bundles}

\resetcounter

Let $\pi:X\ra B$ be a Calabi-Yau space, elliptically fibered over the base 
$B$ (embedded via a section $\si$) and $V$ an $SU(n)$ bundle, 
semistable on the generic elliptic fibre $F$, so $V$ arises by the 
spectral cover construction, cf.~app.~\ref{data} for notations. 
Associated to $V$ is a spectral surface $C$ (an $n$-fold ramified cover
of $B$) of class $n\si+\pi^*\eta$ and let the spectral twist parameter
$\la \in \frac{1}{2}{\bf Z}$
be kept fixed (we assume $h^{1,0}(C)=0$, as $C$ may be ample, say).

Furthermore we consider the case that the (isolated) rational instanton curve 
$b$ lies in the base $B$; its respective normal bundles in $B$ and 
the elliptic surface $\E=\pi^{-1}\, b$ over $b$
have first Chern classes given by the respective self-intersection numbers
$c_1\, (N_{B}b)=b^2$ and $c_1\, (N_{\E}b)=s^2=-\chi$
with $s=\si|_{\E}$ and  $\chi:=c_1\cdot b$
(so $\T:=K_B^{-1}$ becomes $\cO_b(\chi)$ after restriction to $b$), 
cf.~[\ref{extrinsic}]; by restriction one gets the spectral curve $c\subset \E$
over $b$ of class $ns+rF$ where $r=\eta\, b$
(these cohomological data will be held fixed throughout).

The pfaffian factors through the restriction maps 
between the bundle moduli spaces
\beqa
\M_X\, \big(V\big)&\lra & \M_{\E}\, \big( V_{\E}\big) \;\;\;\;  \lra \; 
\M_b\, \big(V_b\big)
\eeqa
The difference between bundles $V_{\E}$ (or $V_b$) {\em defined from 
the outset just over $\E$ (or $b$)} and the restrictions
$V|_{\E}$ (or $V|_b$) of our bundle $V$ (defined over $X$),
i.e. between the full moduli spaces on the right and the image(s) 
from the left, will be important (cf.~sect.~\ref{moduli spaces}).

\subsection{The family of chiral Dirac operators in the elliptic case}

The base space of the Pfaffian line bundle
will be the moduli space $\M_X(V)$ of the spectral bundle $V$
(the curve $b$ is held fixed). 
For bundles defined via the spectral cover construcion the whole
situation will be pulled-back to an situation uplifted over $b$. 
This happens by consideration of a line bundle 
(fixed up to a discrete choice $\la \in \frac{1}{2}{\bf Z}$)
over a cover curve $c$ of $b$;
this curve $c$ varies in accordance with a motion
in the bundle moduli space $\M_X(V)$. 
The considerations of sect.~\ref{section of line bundle}
thus apply now just on the uplifted level.

We fix the parameters $n, \eta, \la$
and consider from now on just the corresponding component $\M_X^{(\la)}(V)$.
We have a family of bundles 
(with $m$ the varying modulus in $\M_X^{(\la)}(V)$)
\beqa
\label{starting family}
\underline{{\cal S}}\;\;\;\;\;\;\;\;\;\;\;\;\;\;\;\;\nonumber\\
\da\; S_b\ox V_m|_b\\
\M_X^{(\la)}(V)\;\;\;\;\;\;\;\;\;\;\nonumber
\eeqa
cf.~(\ref{family of bundles}). The base can be identified with 
the linear system $|C|={\bf P}H^0(X, \cO_X(C))$.

In this preliminary subsection we just want to give a first idea 
how the general approach is expressed in our concrete case.
For spectral bundles one is led
to consider the moduli of the spectral curve $c$ over $b$. 
Then one relates (\ref{starting family}) to the family
\beqa
\label{restricted family}
{\cal S}\;\;\;\;\;\;\;\;\;\;\;\;\nonumber\\
\da\; S_c\ox \cF_c\\
\M_{\E}^{(\la)}(V|_{\E})\;\;\;\;\;\nonumber
\eeqa
The base will be identified with the linear system 
$|c|={\bf P}H^0(\E, \cO_{\E}(c))$ 
as here now the curve $c\subset \E$ varies;
the flat line bundle $\cF$ varies along with the concrete
specific curve $c'=(s)\in |c|$, for $s\in H^0(\E, \cO_{\E}(c))$,
by restricting a universal line bundle on $\E$ to the respective curve
(later, cf.~remark after (\ref{the decomposition}) and 
footnotes \ref{integralities} and \ref{notation adjusted for n odd}, 
we will point to some subtleties pertaining to the question
where precisely which line bundle is defined).

One has with the $n$-fold cover curve $c$ of $b$ 
in the elliptic surface $\E$ 
and the flat line bundle $\cF\in Pic^0(c)$ on $c$ 
(note the corresponding facts $c_1(V|_b)=0$ and $c_1(\cF)=0$) that 
\beqa
S_-(b)\ox V|_b & = & \pi_{c\, *}\; \Big(S_-(c)\ox \cF\Big)
\eeqa
This relation has as a consequence the identification 
\beqa
\label{sections}
\Gamma\Big(b, S_-(b)\ox V_b \Big)& \cong & \Gamma\Big(c,S_-(c)\ox \cF\Big)
\eeqa
(the question of spin structures will be discussed later).
As contribution of $b$ to the superpotential just means 
absence of non-trivial holomorphic sections on the left hand side, one finds 
a precise condition (cf. below) in the moduli of $c$:
the continuous moduli of $V$ stem from
external motions of the spectral surface $C$ in $X$ which in $\E$ map to
$\M_{\E}(c)={\bf P}H^0(\E, \cO_{\E}(c))=|c|$; these external motions
of $c$ in $\E$ are described by the coefficients of its defining polynomial,
and they map also to the intrinsic moduli of $c$
\beqa
{\bf P}H^0\Big(X, \cO_{X}(C)\Big )\; \lra \; 
{\bf P}H^0\Big(\E, \cO_{\E}(c)\Big )& \lra & \M_g
\eeqa
The fixed line bundle $\widetilde{\cL}=l(-F)$ (for a fixed chosen $\lambda$) 
associates to the varying $c'\in |c|$ an element 
$\cL_{c'}=l(-F)|_{c'}\in Pic^0(c')$.
Consideration of the sections in (\ref{sections}) therefore leads, using
the map $\mu: \; Pic^0(c)\; \lra \; Jac(c)$, to the criterion for
contribution to the superpotential
(both sides taken at a specific curve $c'\in \M_{\E}(c)$; 
for details cf.~sect.~\ref{intrinsic})
\beqa
W_b\; \neq \; 0 &\Longleftrightarrow & 
\theta \Big( \mu (\cF) \Big) \; \neq \; 0
\eeqa

\subsection{\label{moduli spaces}Structure of the moduli space
of the bundles over $X$}

Let us recall the structure of the moduli space of spectral bundles. 
Just as in eight dimensions the spectral points on the elliptic curve 
represent the degrees of freedom of the bundle, one expects
by adiabatic extension that moduli arise from the deformations 
of the spectral object in its ambient space; 
the number of the deformations in ${\bf P}H^0(X, \cO_X(C))$
is then $h^{d,0}(C)$ (cf.~below) for a $d$-dimensional spectral object in an 
ambient {\em Calabi-Yau space} as its normal bundle equals then its 
canonical bundle. In studies of $N=2$ string-duality in six dimensions 
that ambient space was a $K3$ surface, elliptically fibered over a ${\bf P^1}$;
the genus of the spectral curve equals then the quaternionic dimension 
of the full quaternionic bundle moduli space, 
given by the total space of the fibration below in (\ref{one-dim situation}).

Consider the corresponding structure of the moduli space
for the bundle $V$ defined over the whole of $X$.
The moduli space of $V$ shows a fibration structure
(for ${\cal A}_B$ cf.~below)
\beqa
\label{fibration structure}
\M_X(V)\;\;\;\;\;\;\;\;\;\;\;\;\nonumber\\
\da \; H^1\big(B, {\cal A}_B\big)\\
{\bf P}H^0\Big(X, \cO_X(C)\Big)\;\;\;\;\nonumber
\eeqa
We consider first only the tangent space to this fibration, i.e. 
first-order deformations\footnote{which are unobstructed [\ref{FMWIII}],
however, for $C$ smooth and $B$ rational}.

Beginning with the base in (\ref{fibration structure})
note that after tensoring the short exact sequence
$0\lra \cO_X(-C)\lra \cO_X\lra \cO_C\lra 0$ with
$\cO_X(C)$ and taking cohomology one gets the 
isomorphism\footnote{as $h^{0,1}(X)=0$; when applying 
the same argument to $\E$ and $c$ therefore $h^{0,1}(\E)=0$ will be relevant}
$T{\bf P}H^0\Big(X, \cO_X(C)\Big)\cong H^0(C, N_C)\cong H^0(C, K_C)$, 
cf.~above. On this tangential level 
the fibration (\ref{fibration structure}) arises
from the Leray spectral sequence 
\beqa
\label{Leray}
H^1(X, ad\, V)\;\;\;\;\;\;\;\;\;\;\;\;\;\;\;\;\;\;\;\;\;\;\;\;\;\;\;\;
\;\;\;\;\;\;\;\;\;\;\;\;\;\;\;\;\nonumber\\
\da \; H^1(B, \pi_*\, ad\, V)\cong H^1(C, \cO_C)\;\;\;\;\;\;\;\;\\
\;\;\;\;\;\;\;\;
H^0(B, R^1\, \pi_*\, ad\, V)\;\;\;\;
\cong H^0(C, K_C)\;\;\;\;\;\;\;\;\;\;\nonumber
\eeqa
(using the identification of the abstract space
$\pi_*\, ad\, V$ with $\pi_{C*}\, \cO_C$ and, 
concerning the space of deformations 
of $C$ in $X$, the identification of $R^1\, \pi_*\, ad\, V\;$ 
with\footnote{cf.~for (\ref{Leray}) at this stage the tangential map 
of (\ref{moduli space fibration map}) and the dimension computation 
(\ref{count of par's}) from coefficients of the defining equation of $C$}
$\pi_*\; \cO_X(C)\; \cong \;
\oplus_{\stackrel{i=0}{i\neq 1}}^n \; \M\otimes \T^{-i}$ 
which in turn
is identified with  $\pi_{C*}\, \cO_X(C)|_C\; \cong \; \pi_{C*}\, K_C$).

Let us go beyond the consideration local in the fibre made here and describe 
how the other components of the full fibre $\{\gamma\}\cong{\bf Z}$ 
(parametrised by $\la$) enter the picture, cf.~[\ref{CD}]. 

So let us discuss the twisting data which 
occur when piecing together along $B$ bundles over the fibers.
Let ${\cal M}_F$ be the moduli space of semistable $SU(n)$-bundles over $F$, 
${\cal M}_{X/B}$ the relative object (fibered over $B$) and $\Xi$ the 
universal bundle over\footnote{the superscript $0$ denotes 
the smooth locus where a universal object exists locally}
$F\times {\cal M}_F^0$ or $X\times _B {\cal M}_{X/B}^0$. 
The  bundle $V$ over $X$, 
fiberwise\footnote{over the open subset of $B$ over which lie 
smooth fibers; the mentioned section is also over that subset} semistable, 
gives a section $\tilde{s}$ of ${\cal M}_{X/B}^0 \rightarrow B$.
Conversely one tries to build $V$ from 
$\tilde{s}$ by pulling back a universal bundle. 

On the fibre there is, associated to $\Xi$, an abelian
group scheme of automorphism groups $\underline{Aut} (\Xi)$ over 
${\cal M}_F^0$ (of associated sheaf of sections ${\cal A}$). The
set of universal bundles over $F\times {\cal M}_F^0$ 
is\footnote{if the obstruction in $H^2({\cal M}_F^0,{\cal A})$ vanishes}
rotated through under $H^1({\cal M}_F^0,{\cal A})$,
or $H^1(B,{\cal A}_B(\tilde{s}))$ in the relative version.

The twisting data $H^1(B, {\cal A}_B)$ are fibered over
its discrete part $\gamma {\bf Z}$ by the 
continuous relative jacobian $Jac(C/B)$, cf.~[\ref{FMWII}].
This is itself the relative version of the fibration
of the Picard group (the moduli space of line bundles) over its discrete part 
(characterised by the first Chern class) by the set of flat bundles
(parametrized  by a complex torus)
\begin{center}
\begin{tabular}{ccccccccc} 
$0$ & $\rightarrow$ & $H^1(B, {\cal A}_B)$ & $\rightarrow$ & $Pic \, C$ & 
$\rightarrow$ & $Pic \, B$ & $\rightarrow$ & $0$ \\
& & & & & & & & \\
 & & $\;\;\;\;\;\;\;\;\;\;\;\;\; 
\;\;\;\; 
\downarrow \; Jac(C/B)$ & & $\;\;\;\;\;\;\;\;\;\;\; \downarrow \; Pic_0 \, C$ 
& & $\;\;\;\;\;\;\;\;\;\; \downarrow \; Pic_0 \, B$ & &  \\
& & & & & & & & \\
$0$ & $\rightarrow$ & $\{\gamma\}\cong {\bf Z}$ & $\rightarrow$ & 
$H^{1,1}(C,{\bf Z})$ &  $\rightarrow$ & $H^{1,1}(B,{\bf Z})$ & 
$\rightarrow$ & $0$ \\
\end{tabular}
\end{center}
The $B$ considered here have trivial $Pic_0(B)$.
As $\pi_{C\, *}(c_1(L))$ is fixed by the condition $c_1(V)=0$
the possible $L$'s are 
parametrized\footnote{\label{Pic(C)}we restrict us to the case 
that $Pic(C)$ is
generated by the restrictions of elements of $Pic(X)$}
by a discrete part encoded in $\la$ and a continuous part from 
$Jac(C/B)\cong Pic_0(C)\cong Jac(C)$.
In case $C$ is ample (positive) one has $\pi_1(C)\cong 1$ such
that $h^{1,0}(C)=0$, {\em as we assume throughout}. 
So over $X$ no continuous moduli appear
as $Pic_0(C)\cong Jac(C)$ will not enter the story,
only the discrete choice in $\ker \, \pi_{C\, *}:\, H^{1,1}(C)\ra 
H^{1,1}(B)$ remains. For simplificity we assume the generic 
situation$^{\ref{Pic(C)}}$ 
that the only classes in this kernel are those visible 
already on $X$, i.e.~arise by restriction; then the remaining discrete part 
is parametrized by a number $\la\in \frac{1}{2}{\bf Z}$
(cf.~app.~\ref{data}).

So, given our assumptions, to let vary a bundle $V$ over its moduli space
$\M_X(V)$ comes down 
(once the discrete parameter $\la\in \frac{1}{2}{\bf Z}$ is fixed) 
to let vary the concrete surface given by 
the zero set $(\underline{s})$  
in the linear system $|C|={\bf P}H^0(X, \cO_X(C))$
(for $\underline{s}\in H^0(X, \cO_X(C))$).
Let denote $\underline{S}_{\la}$ 
the corresponding section of the fibration (\ref{fibration structure})
(whose fibre is just $\{\gamma\}\cong {\bf Z}$)
\beqa
\underline{S}_{\la} \; & \in & \;
\Gamma\Big( {\bf P}H^0\big( X, \cO_X(C)\big), \M_X(V) \Big)
\eeqa

\section{\label{restricted bundles}
The moduli space of the restricted bundles}

\resetcounter

Recall that we consider the following restriction map between 
the moduli spaces
\beqa
\label{restriction of moduli spaces}
\M_X(V) & \lra & \M_{\E}(V_{\E})
\eeqa
Before considering the {\em image of this map} 
(the part of the right hand side relevant for us) in more detail, we note 
that the full moduli space $\M_{\E}(V_{\E})$ of spectral $SU(n)$ bundles 
{\em which are defined from the outset just over $\E$} 
possesses the fibration (with $Jac(c_t)\cong Pic_0(c_t)$)
\beqa
\label{one-dim situation}
\M_{\E}(V_{\E})\;\;\;\;\;\; \nonumber\\
\;\;\;\; \da \; Jac( \cdot ) \\
{\bf P}H^0\Big(\E, \cO_{\E}(c)\Big) \nonumber
\eeqa
(in the base a section $s\neq 0$ of ${\cal O}_{\E}(c)$ defines
a curve given by its zero set $(s)$).

Note that, 
unlike the situation for a two-dimensional base $B$ considered above,
here $\E$ will not be a Calabi-Yau ambient space for the spectral object $c$
(as $\E$ will not be a $K3$ surface in general). The former $\T=K_B^{-1}$, the
line bundle  relevant to the description of the elliptic fibration, becomes 
inside $\E$ now $\T|_b=\cO_b(\chi)=\cO_b(1)$ as $\chi=1$ for $b$ isolated.

By the restriction map $V\, \lra \, V|_{\E}$ the following phenomenon occurs:
the moduli space of bundles over elliptically fibered spaces shows a 
significant difference between the cases where the base is either $B$ or $b$. 
Whereas in the spectral {\em surface} case the continuous part of the 
twist data (the fibre of the moduli space) 
vanishes under generic conditions (like $C$ being ample) 
this is not the case for 
a spectral {\em curve} which comes always naturally equipped with its Jacobian
(but here no discrete twist arises). 
This has the following consequence: a bundle $V$ over $X$, considered to vary 
over its moduli space, is described just by a {\em constant} section 
$\underline{S}_{\la}$ of its fibration (\ref{fibration structure}). 
However, its restriction to $\E$
\beqa
\label{transfer}
\M_{X}(V_X) \;\;\;\;\;\;\;\;\;\;\;
\lra \;\;\;\;\;\;\;
\M_{\E}(V_{\E}) \;\;\;\;\;\;\;\;\;\nonumber\\
\da \; \{\gamma\}\cong {\bf Z}\;\;\;\;\;\;\;\;\;\;\;\;\;\;\;\;\;\;\;\;\;\;\;\;
\da \; Jac(\cdot)\;\;\\
{\bf P}H^0\Big(X, \cO_{X}(C)\Big)\;\;\;\;
\lra \;\;\;\;
{\bf P}H^0\Big(\E, \cO_{\E}(c)\Big)\nonumber
\eeqa
is given by a section $S_{\la}$ of a moduli space fibration
(\ref{one-dim situation}) with the fibre no longer discrete
\beqa
\label{transfer 1}
\Gamma\Big( {\bf P}H^0\big( X, \cO_X(C)\big), \M_X(V) \Big)
\ni \underline{S}_{\la} & \lra & S_{\la}\in 
\Gamma\Big( {\bf P}H^0\big( \E, \cO_{\E}(c)\big), \M_{\E}(V_{\E})\Big)
\;\;\;\;\;\;\;\;
\eeqa
So, for a bundle {\em defined a priori just over $\E$} the total space 
$\M_{\E}(V_{\E})$ in the fibration (\ref{one-dim situation}) is relevant 
as describing a varying object; for our case, with a $V_{\E}$ 
{\em arising as restriction $V_{X}|_{\E}$ 
where $V_{(X)}$ is defined already over $X$}, 
the image of the restriction map (\ref{restriction of moduli spaces}) 
rather is (holding fixed $\la \in\frac{1}{2}{\bf Z}$, 
giving a trivial fibre in (\ref{fibration structure})) 
just one section $S_{\la}$ of (\ref{one-dim situation}).

\subsection{Decomposition of $\cL$ relative to a spin choice}

The moduli of $V$, considered over $X$, consist, besides the fixed 
discrete choice $\la \in \frac{1}{2}{\bf Z}$, in the motions of $C$
in $X$, i.e.~they are given (up to an overall rescaling) 
by the coefficients of the sections $a_i$ in the defining polynomial, 
cf.~app.~\ref{spectral equation}; 
the world-sheet instanton superpotential $W_b$ will depend only on their
restrictions to $b$. For $B$ the Hirzebruch surface ${\bf F_k}$, say,
the $a_i$ are certain polynomials $a_i(z_1, z_2)$ 
(whose coefficients are essentially the moduli)
and for $W_b$ only the coefficients are concerned
which remain after setting $z_2=0$ (i.e.~after restricting from $B$ to $b$);
in other words, $W_b$ depends effectively 
only on the image of a section to the restriction map of 
$\M_X(V)$ on its image in $\M_{\E}(V_{\E})$. 

Before proceeding further let us keep on record here two important
line bundles 
\beqa
\Lambda&=&\cO_{\E}\Big(ns-(r-n)F\Big)\\
K_c    &=&\cO_{\E}\Big(ns+(r-1)F\Big)\Big|_c
\eeqa
The importance of $\Lambda$ stems from the fact 
that\footnote{for $gcd(n, r-n)=1$, which is fulfilled automatically
in the cases of application (where $n\leq 5$)}
it generates, among the bundles $\cO_{\E}(ps+qF)$ 
on $\E$ (i.e.~the ones which come from $X$), those 
which become flat on $c\simeq ns+rF$.

For a spectral cover bundle $V=p_*(p_C^*L\ox \cP)$ over $X$,
with $L=\un{L}|_C$ a line bundle on $C$ arising as restriction from $X$, 
one has for $\widetilde{\cL}=l(-F)$ with $l=\un{L}|_{\E}$ 
\beqa
\label{l(-F) on E}
\widetilde{\cL}=\left\{ \begin{array}{ll}
\cO_{\E}\Big(\frac{1}{2}\Big[ns+(r-1)F\Big]\Big)\ox
\cO_{\E}\Big(\la\Big[ns-(r-n)F\Big]\Big)
& \mbox{for} \; n \, \mbox{even} \; 
(\ra \la \in {\bf Z}, r\, \mbox{odd}) \\
\cO_{\E}\Big((\la+\frac{1}{2})ns
+\Big[(\la n - \frac{1}{2})-(\la-\frac{1}{2})r\Big]F\Big)
& \mbox{for} \; n \, \mbox{odd}\; (\ra \la \in \frac{1}{2}+{\bf Z})
\end{array} \right.
\eeqa
(the implications for $n$ even/odd follow by the remark
after (\ref{alpha beta definitions}); so the expressions for
$\widetilde{\cL}$ exist).
For us $V_{\E}$ is $V_{(X)}|_{\E}$ with $\widetilde{\cL}|_c=\cL$ 
a flat twist of a theta characteristic 
\beqa
\label{the decomposition}
\cL&=&K_c^{1/2}\ox \cF
\eeqa
Here for $n$ even, cf.~(\ref{l(-F) on E}), we use the standard representative 
$\cO_{\E}(\frac{1}{2}[ns+(r-1)F])|_c$ for $K_c^{1/2}$
coming already from $\E$ (canonical "global spin choice", here 'global'
refers to the variation of $c$ in $\M_{\E}(c)$); as $\la\in {\bf Z}$
also $\cF=\Lambda|_c^{\la}=\cO_c(G|_c)$ exists individually.
For $n$ odd, considered {\em on $\E$}, only the combination in 
(\ref{l(-F) on E}) exists; however, $c$ being a curve, a square root of the 
canonical bundle will always exist, though having a $Pic_0^2(c)$ ambiguity; 
we will consider a globalization over $\M_{\E}(c)$ of such a choice 
as part of the data (a non-unique "global spin choice"),
keeping in mind its indicated non-uniqueness.
A flat bundle $\cF_c$ will then also exist 
(as $\cL$ exists independent of choices), depending on the choice indicated.
Then, once a global spin choice is made, in 
$\cF_{c_t}$ it is the variation inside $\E$ of the respective 
concrete curve $c_t$ in the linear system
$|c|={\bf P}H^0(\E, \cO_{\E}(c))=\M_{\E}(c)$, which gives the 
respective position of $\cF_{c_t}$ in a Jacobian $Jac(c_t)$.

\subsection{Dimension of the moduli space and the divisor $(Pfaff)$}

For $c$ positive one finds by the index theorem 
(the second relation holds in general)
\beqa
h^0\Big(\E, \cO_{\E}(c)\Big)& = & n(r+1)-\Big(\frac{n(n+1)}{2}-1\Big)\chi\\
\deg \, l(-F)|_c \; = \; 
\deg  K_c^{1/2} \; & = & n( r - 1) -\Big( \frac{n(n-1)}{2}\Big)\, \chi
\eeqa
The moduli space of external motions of $c$ has dimension
$h^0\big(\E, \cO_{\E}(c)\big)-1$ and for $b$ isolated we have $\chi=1$ 
[\ref{extrinsic}]
such that one gets then (where $\dim  Jac(c) = \deg\; K_c^{1/2} +1 $)
\beqa
\label{dim of restricted}
\dim \M_{\E}(c)\; & = & n\Big( r - \frac{n-1}{2}\Big)\;\;\;\;\;\;\;\;
=\, g-1+n\\
\label{genus}
\dim  Jac(c)   \; & = & n\Big( r - \frac{n+1}{2}\Big) +1\;\; = \, g\\
\dim \M_{g_c}  \; & = &3n\Big( r - \frac{n+1}{2}\Big)\;\;\;\;\;\;\, =\,  3g-3
\eeqa
A concrete basis $\omega_{\alpha}$, $\alpha = 1, \dots, g$, of $H^{1,0}(c_t)$ 
arises by taking Poincare residues of elements in
$\Omega^2_{\E}(c)$, written locally as $\omega=\frac{h}{w}du\wedge dX$
where $u=u_1/u_2$ and $X=x/z$ are affine coordinates on $b$ and $F$, 
resp.~(on $\E-s-F$, fibered by $F-\{p_0\}$ over
$b-\{q_0\}$) and $w=0$ the spectral curve equation. 
Because $K_{\E}=\cO_{\E}(-F)$ one gets 
$h=h_0+h_2x+h_3y+\dots +h_nx^{n/2}\in H^0(\E, \cO_{\E}(ns+(r-1)F))$ 
globally\footnote{for $n$ even, say, cf.~app.~\ref{coordinates}; the 
coefficient functions $h_i(u_1, u_2)$ have degree $r-1-i$; 
the $h_{\al}$ with $\al=1, \dots, g=d(n, r-1)$ are suitably enumerated 
also by $h_{\al}=h_{ij}$ with $i=0,2,3, \dots, n, j=0, 1, \dots, r-1-i$}
such that $\dim \M_{\E}(c)=d(n, r)-1$ and $g=d(n, r-1)$ where
$d(n, r):=h^0(\E, \cO_{\E}(c))$; concretely one gets
$\omega_{\al}=\frac{h_{\al}}{\partial_u w}dX=-\frac{h_{\al}}{\partial_X w}du$.

The fact that $\dim \M_{\E}(c)\neq \dim  Jac(c)$
reflects the fact that $\E$ is not a $K3$ surface, 
so no hyperk\"ahler structure for $\M_{\E}(V_{\E})$ arises
(the equality of dimensions is recovered, however, for the Calabi-Yau space 
over ${\bf F_0}$ where $\E$ becomes $K3$ and $\chi=2$ gives $n(r-n)+1$).

The fact that $\dim \M_{\E}(c) < \dim  \M_g$ reflects the fact that
the curves $c$ of class $ns+rF$ and of genus $g_c$ arising in $\E$
are not the most general curves of genus $g$ (cf.~the non-genericity
of plane curves in ${\bf P^2}$). For example, because of
the $n$-fold covering of $b$, 
they carry a certain $r$-dimensional system $g_d^r$ of divisors of degree $d$, 
here a $g_n^1$; however the generic genus $g$ curve has such a system 
only if $g\leq 2n-2$, what contradicts the genus (\ref{genus}) 
(we always have $r>n$, cf.~app.~\ref{data}). The space $\M_{g, n}^1$
of curves having a $g_n^1$ has dimension $2g+2n-5$ if, as in our case,
$n\leq \frac{g+2}{2}$ (note, for comparison, $\dim \M_g^{hyp}=2g-1$).

The degree of the Pfaffian polynomial $Pfaff=\det \; \iota_1$
is given by 
$h^1(\E, l(-F)) = (\la^2-\frac{1}{4})n(r-\frac{n}{2})-1$.
Therefore the $W_b=0$ locus is given by a divisor
$(Pfaff) \subset  {\bf P}H^0\big(\E, \cO_{\E}(c)\big)$
of degree $2k-n-1$ (for $\la=3/2$, say)
inside a ${\bf P}^{k}$ where $k=n(r - \frac{n-1}{2})$.

\subsection{\label{loci}Special loci in the moduli space: 
$\Sigma \subset (Pfaff)$}

Recall that $\cL=K_c^{1/2}\ox \Lambda|_c^{\la}$ 
(cf.~(\ref{l(-F) on E}), remarks there and footn.~\ref{integralities})
where $\la \in \frac{1}{2}{\bf Z}$ and
\beqa
\Lambda&=&\cO_{\E}\Big(ns-(r-n)F\Big)
\;\;\;\;\;\;\;\;\;\;\;\;\;\;\;\; \mbox{with}\;\;\;
\Lambda|_{c_t}\in Pic_0(c_t)
\eeqa
Define the subset $\Sigma_{k}$ of $\M_{\E}(c)$ where $\Lambda|_{c_t}$ 
becomes a $k$-torsion element. So (cf.~app.~\ref{theta characteristics})
\beqa
\label{Sigma_lambda}
\Sigma_{2\la}&=&
\Big\{\, t\in \M_{\E}(c)\,  \Big| \; \Lambda|_{c_t}\in Pic_0^{2\la}(c_t)\;\;
\Big(\Llra \; \cL_{c_t}\in  \cS p(c_t)\Big)\Big\}\\
\Sigma&=&
\Big\{\, t\in \M_{\E}(c)\,  \Big| \; \Lambda|_{c_t}\cong \cO_{c_t}\Big\}
\; \subset \; \Sigma_{2\la}
\eeqa
(for the definition of $\Sigma=\Sigma_{1}$ 
cf.~sect.~3.2.1 of [\ref{extrinsic}]).
$\Sigma_{2\la}$ 
is where $\cL^2_{c_t}\cong K_{c_t}$ holds; this is just the locus where
$\tau^*V|_{\E}\cong V^*|_{\E}$ [\ref{FMW}] 
(for $\tau$ cf.~app.~\ref{dimension}), such that, by $(\tau^*V)|_B=V|_B$,
\beqa
\label{Sigma remark 1}
t\in \Sigma_{2\la}&\Lra & \;\; V|_b\cong V^*|_b\;\;\;\;\;\;\;\;\;\;\;\;\\
\label{Sigma remark 2}
t\in \Sigma \;\, &\Lra &Pfaff(t)=0
\;\;\;\;\;\;\;\;\;\;\;\;
\eeqa
Here (\ref{Sigma remark 2}) is obvious 
as the effectivity of $D$ in $\cL=\cO_c(D)$ is sufficient
for $Pfaff$ to vanish (cf.~also (\ref{corol})):
 now one gets\footnote{\label{integralities}actually 
one gets the same conclusion for $\la\in \frac{1}{2}+{\bf Z}$ 
as in the case of $\la \in {\bf Z}$, despite first appearence:
note that not only has $\frac{1}{2}K_{c_t}$ 
the {\em integral} degree $n(r-\frac{n+1}{2})$\,\, - \, 
clearly a necessary condition for the existence 
of the bundle $\cO_{c}(\frac{1}{2}(ns|_c+(r-1)F|_c))$\, - \,
but, $c_t$ being a curve, the bundle $K_{c_t}^{1/2}$ will always exist; 
therefore $\Lambda|_{c_t}^{\la}$ will exist as well because the combination
$\cL_{c_t}=K_{c_t}^{1/2}\ox \Lambda|_{c_t}^{\la}$ exists (this
because $\cL_{c_t}$ is the restriction to $c_t$ of
the line bundle $\widetilde{\cL}$ which exists already on $\E$, 
by our standing integrality assumptions, cf.~app.~\ref{data}); now, however, 
we do not know that $\Lambda|_{c_t}^{\la}\cong \cO_{c_t}$ but rather only 
that $\Lambda|_{c_t}^{\la}\in Pic_0^2(c_t)$ 
(actually this expression is not even defined more precisely because
similarly we do not know precisely which square root of $K_{c_t}$ here
the $K_{c_t}^{1/2}$ is, as it will not arise by restriction from $\E$
as was the case for $\la$ integral; that is, here the individual factors
would depend essentially on a spin choice made);
so $\cL$ becomes now (instead of the explicit $K_{c_t}^{1/2}$ 
in (\ref{cL specialisation}) for $\la\in {\bf Z}$) 
for $\la\in \frac{1}{2}+{\bf Z}$ a different theta characteristic 
(spin bundle structure, i.e.~square root of $K_{c_t}$, 
cf.~app.~\ref{theta characteristics})
$K_{c_t}^{1/2}\ox \Lambda|_{c_t}^{\la}$ which means here concretely 
the second line of (\ref{l(-F) on E}) restricted to $c$; 
then the argument can be completed as indicated in (\ref{cL specialisation}),
with the ultimate result being independent of any potential spin choices} 
along $\Sigma$ that (for the parities cf.~(\ref{alpha beta definitions}))
\beqa
\label{cL specialisation}
\cL\stackrel{\Sigma}{\ra}\left\{ \begin{array}{ll}
K_{c_t}^{1/2}=\cO_{\E}\Big(\frac{n}{2}s+\frac{r-1}{2}F\Big)\Big|_{c_t}
& \mbox{for} \; \la \in {\bf Z} \; 
(\ra n\, \mbox{even}, r\, \mbox{odd}) \\
\cO_{\E}\Big(\Big[(\la+\frac{1}{2})(r-n)+\beta\Big]F\Big)\Big|_{c_t}
=\cO_{\E}\Big(\Big[r-\frac{n+1}{2}\Big]F\Big)\Big|_{c_t}
& \mbox{for} \; \la \in \frac{1}{2}+{\bf Z} \; (\ra n\, \mbox{odd})
\end{array} \right.
\eeqa
So we have the array of inclusions
(we assume we are not in a case where $Pfaff\equiv 0$)
\beqa
\begin{array}{ccccccc}
       &      &                                &         &(Pfaff) &&\\
       &      &                                &\nearrow & & \searrow & \\
\Sigma & \lra & \; \Sigma_{2\la}\cap (Pfaff)\; & & & & \M_{\E}(c)\\
       &      &                                &\searrow & & \nearrow & \\
       &      &                                &         &\Sigma_{2\la}&&
\end{array}
\eeqa
where $(Pfaff)\hookrightarrow \M_{\E}(c)$ has codimension one
(for the restriction to $\Sigma_{2\la}$ cf.~Rem.~$(2)$).

\noindent
{\em \un{Remarks:}} \hspace{.2cm} Let us add some remarks on a 
type of set like $\Sigma_{2\la}$.\\
(1) $\Sigma_{2\la}\subset \M_{\E}(c)$ is the locus where
$\Lambda|_{c_t}^{\la}\in Pic_0^2(c_t)$; here $\Lambda|_{c_t}^{\la}$ is for 
$n$ odd only {\em defined} up to $Pic_0^2(c_t)$ as then also the factor 
$K_c^{1/2}$, to be split off from $\cL_{c_t}$, is defined 
only up to that ambiguity (here $\la \in \frac{1}{2}{\bf Z}$ whereas 
above we had $k\in {\bf Z}$; cf.~footn.~\ref{integralities}).

(2) Note that $\Sigma_{2\la}$ decomposes 
as $\Sigma_{2\la}^+\stackrel{\cdot}{\cup} \Sigma_{2\la}^-$
according to the parity of $\cL_{c_t}$ (that is of $h^0(c_t, \cL_{c_t})$)
such that one gets
\beqa
\Sigma_{2\la}\cap (Pfaff)&=&\Big(\Sigma_{2\la}^+\cap (Pfaff)\Big)
\;\stackrel{\cdot}{\cup} \; \Sigma_{2\la}^-
\eeqa
This might be compared to (\ref{Z Theta decomposition}); accordingly one has 
under the period map in (\ref{one-dim situation with abstract fibration})
\beqa
\Sigma_{2\la}^+\cap (Pfaff)&\stackrel{\Pi}{\lra}&\M_g^1\; \subset \, \M_g
\eeqa
So it is the component $\Sigma_{2\la}^+\cap (Pfaff)$ which has codimension one
in $\Sigma_{2\la}^+$ (or $\Sigma_{2\la})$.

(3) The obvious relation $\Sigma \subset \Sigma_{2\la}$ 
is a special case of the general fact
$\Sigma_{k'}\subset \Sigma_{k}$ for $k'|k$. Note, however, that
here the codimension of $\Sigma_{k'}$ in $\Sigma_{k}$ cannot be $>0$:
in that case one would have a continuous deformation from elements
of $\Sigma_{k}-\Sigma_{k'}$ to elements of $\Sigma_{k'}$ 
which is impossible (as elements of strictly higher torsion order cannot be 
deformed, inside $\Sigma_{k}$, to elements of lower torsion order); 
so $\Sigma_{k}$ is rather reducible with $\Sigma_{k'}$ a component. 
So $\Sigma$ is a component of $\Sigma_{2\la}$; note also that 
$\Sigma_{k}\cap \Sigma_{k'}=\Sigma_{gcd(k,k')}$;
$\Sigma_{2\la}$ has components 
\beqa
\label{component decomposition}
\Sigma_{2\la}&=&{\displaystyle \stackrel{\cdot}{\bigcup}_{k'|2\la}} 
\; \Sigma_{k'}^{strict}\;\;\;\;\;\;\;\;\;\;\;\;\; 
\eeqa
comprising elements of $\Sigma_{k'}$ of {\em strict} torsion order $k'$; 
for $p$ prime, say, one has 
$\Sigma_{p}=\Sigma\stackrel{\cdot}{\cup}\Sigma_{p}^{strict}$
where $\Sigma\subset (Pfaff)$; compare the relation 
$\Sigma_{p}=\Sigma_{p}^+\stackrel{\cdot}{\cup} \Sigma_{p}^-$
with $\Sigma_{p}^-\subset (Pfaff)$.

(4) Besides $\Sigma$ one has, via the same argument,
further components lying in $(Pfaff)$
\beqa
\label{stronger even}
\Sigma_{\la}
&\subset& (Pfaff)\;\;\;\;\;\;\;\;\;\;\;\;\;\;\;\;\;\; \;\;\;\;\;\; 
\mbox{for}\; n\equiv 0 \, (2) \\
\label{stronger odd}
\Sigma_{\la+\frac{1}{2}}
&\subset& (Pfaff)
\;\;\;\;\;\;\;\;\;\;\;\;\;\;\;\;\;\; \;\;\;\;\;\; 
\mbox{for}\; n\not\equiv 0 \, (2) 
\eeqa
This follows as the specialisations in (\ref{cL specialisation}) 
still hold in these larger loci.
On a further locus 
$\cL$ becomes $\cO_{\E}(ns+\frac{n-1}{2}F)|_{c_t}$, 
again the line bundle of an effective divisor,
\beqa
\Sigma_{\la-\frac{1}{2}}
&\subset& (Pfaff)
\;\;\;\;\;\;\;\;\;\;\;\;\;\;\;\;\;\; \;\;\;\;\;\; 
\mbox{for}\; n\not\equiv 0 \, (2) 
\eeqa

There are also special cases 
like $\Sigma_{\la - \frac{3}{2}}\subset (Pfaff)$ 
for $n=3,r=4$ and $n=5, r=6,7$. A further special case 
is $\la=9/2$ with $\Sigma_2\subset (Pfaff)$ for $n=5,r=6$.

\subsection{\label{coordinates}
A further explicit locus in the moduli space: $\R\subset \Sigma$}

We will denote points of $\E\subset \cW_{b}$ by pairs $\{p, u\}$ where 
$p$ and $u$ are coordinates of ${\bf P^2}$ and $b\cong {\bf P^1}$, 
respectively. One gets the divisors (with $p_0=(0,1,0)$ and $a_n(u_j)=0$)
\beqa
s|_c&=&\sum_{j=1}^{r-n}\{p_0, u_j\}\\
F_u|_c&=&\sum_{i=1}^n \{q_i, u\}
\eeqa
Let $a_n=\prod_{j=1}^{r-n}a_n^{(j)}$ be a decomposition in linear factors
with $a_n^{(j)}(u_j)=0$ and let 
\beqa
R_i^{(j)}:=Res(a_i, a_n^{(j)})
\eeqa
denote the resultant (cf.~app.~I.1 of [\ref{extrinsic}])
for $i=2, \dots, n-1$ and $n>2$.
If (cf.~sect.~3.2.2 of [\ref{extrinsic}])
$R_i^{(j)}=0$, i.e.~$a_n^{(j)}|a_i$, for all $i=2, \dots, n-1$, 
{\em for one specific} $j$ then
\beqa
ns|_c&=&\sum_{k=1}^{r-n}\{np_0, u_k\}\; = \; 
F_{u_j}|_c+\sum_{\stackrel{\scriptstyle k=1}{k\neq j}}^{r-n}\{np_0, u_k\}
\eeqa
as then $q\in F_{u_j}|_c$ is forced to have $z=0$, i.e.~to be $p_0$
(here $p_0$ is already among the $n$ points $\{q_i^{(j)}, u_j\}$, say
for $i=1$; then the points for $i=2, \dots, n-1$ are forced to be at $p_0$
and the final one follows to be there as well because the points sum up to
zero, represented by $p_0$, in the group law). Thus $(ns-F_{u_j})|_c=$ 
effective and so $(ns-F)|_c\sim$ effective. 
Posing $n-2$ conditions leads to a locus 
of codimension $n-2$. If, however, $R_i^{(j)}=0$ for all 
$i=2, \dots, n-1$ and even {\em for all} $j=1, \dots, r-n$ then one has
(cf.~(\ref{G definitions}))
\beqa
\label{twist class triviality}
ns|_c &=& \sum_{j=1}^{r-n}F_{u_j}\, \sim \, 
(r-n)F|_c\;\;\;\;\;\;\;\;\; \Longrightarrow \; \;\;\; G|_c \;\sim \; 0
\eeqa
or equivalently $\Lambda|_c=\cO_{\E}(G)|_c\cong \cO_c$.
As posing $(n-2)(r-n)$ conditions is sufficient for $G|_c\sim 0$ 
the latter will hold at a locus of codimension $\leq (n-2)(r-n)$,
cf.~also (\ref{Sigma remark 3}).
So one finds
(where as usual 
$(R_i^{(j)})$ denotes the (vanishing) divisor of the polynomial $R_i^{(j)}$)
\beqa
\label{Sigma relation}
t\in \Sigma \; \Llra \; 
\Big(ns-(r-n)F\Big)\Big|_{c_t} \sim \mbox{effective}& \Llra&
\sum_{j=1}^{r-n}\sum_{i=1}^n\{p_0-q_i^{(j)}, u_j\}_{c_t} \sim\mbox{effective}
\;\;\;\;\;\;\;\;\;\;\;\;\\
\label{curly R relation}
t\in \fbox{\mbox{$\R:={\displaystyle 
\bigcap_{\stackrel{\scriptstyle i=2, \dots, n-1}{j=1, \dots r-n}}}\,
(R_i^{(j)})$}} &\Lra& 
\sum_{j=1}^{r-n}\sum_{i=1}^n\{p_0-q_i^{(j)}, u_j\}_{c_t}=\mbox{effective}
\;\;\;\;\;\;\;\;\;\;\;\;
\eeqa
where the effective divisor in the second line is of course the 
zero divisor (i.e.~$F_{u_j}|_c=\{np_0,u_j\}$ for all $j$).
So the locus $\R\subset \M_{\E}(c)$, 
defined explicitly by the equations $R_i^{(j)}=0$, 
is a sublocus of the structurally defined locus $\Sigma$,
cf.~(\ref{Sigma remark 3}) below; cf.~also app.~\ref{aux locus}.

Above we did some general considerations in $Pic(c)\cong Div(c)/\sim$ 
leading to the relation $\R\subset \Sigma$. On the other hand
(\ref{Sigma remark 2}) gave some understanding of non-topological 
(codim $>0$) vanishing loci of $Pfaff$: $\Sigma$ is contained in 
the codim $1$ locus given by the vanishing divisor $(Pfaff)$ (in 
[\ref{extrinsic}] we looked for whole components of this reducible divisor). 
Although the {\em structurally} defined $\Sigma$ is not easily described 
{\em explicitly} in $\M_{\E}(c)$ 
we now have identified here a concretely described sublocus: the locus $\R$
\beqa
\label{Sigma remark 3}
\begin{array}{|c|}
\hline
\; \R \;\; \subset  \;\; \Sigma \;\;\; \subset \;\; (Pfaff)\;\\
\hline
\end{array}
\eeqa
which is of codimension $(n-2)(r-n)$ in $\M_{\E}(c)$. 
Then $\dim \Sigma \geq  \dim \R=n\frac{n-3}{2}+2r$ from (\ref{Sigma remark 3}) 
gives in the first few cases (where $r>n$, cf.~app.~\ref{data})
\beqa
\begin{array}{|c|c||c|c|c|c|c|}
\hline
n & g &\dim\Sigma\geq \dim \R =
&\dim\M_{\E}(c)&\dim\M_{g, n}^1&\dim\M_g^{hyp}&\dim\M_g \\
\hline
2 & 2r-2  & 2r-1  & 2r-1   & 4r-5    & 4r-5   & 6r-9    \\
3 & 3r-5  & 2r    & 3r-3   & 6r-9    & 6r-11  & 9r-18   \\
4 & 4r-9  & 2r+2  & 4r-6   & 8r-15   & 8r-19  & 12r-30  \\
5 & 5r-14 & 2r+5  & 5r-10  & 10r-23  & 10r-29 & 15r-45  \\
\hline
\end{array}
\eeqa

Above we considered the relation of the structurally defined locus
$\Sigma$ and the explicitly described locus $\R$ in general. Let us 
now specialise to some important low ranks: first to the somewhat
special case of $SU(2)$ bundles and then, below in 
sect.~\ref{The SU(3) case}, to our main example class, 
the $SU(3)$ bundles.

\noindent
{\em \fbox{$SU(2)$ bundles} $\; \Sigma=\M_{\E}(c)$, i.e.~$Pfaff\equiv 0$}

Here one has $a_2(u_j)=0\Lra z=0$, cf.~(\ref{explicit coordinate forms}),
such that $F_{u_j}|_c=\{2p_0, u_j\}$ and so $\Sigma=\M_{\E}(c)$
\beqa
\label{n=2}
n=2 & \Longrightarrow & G|_c\; \sim \; 0 , \; i.e.~\Lambda|_c\cong \cO_c
\eeqa

Here, for $SU(2)$ bundles, $c$ is hyperelliptic 
(note also that $\M_g^{hyp}=\M_{g, 2}^1$)
and one has $Pfaff\equiv 0$ as $\Sigma=\M_{\E}(c)$, cf.~(\ref{n=2})
\beqa
\Sigma=\M_{\E}(c) & \Lra & Pfaff\equiv 0
\eeqa

Before we go on in sect.~(\ref{The SU(3) case}) to the more specialised 
considerations in the case of $SU(3)$ bundles, which constitute our main class 
of examples, we insert in the next subsection some
discussion on the question of multiplicity of a factor in the Pfaffian.

\subsection{\label{multiplicities}Some remarks about multiplicities}

We are interested in the multiplicity of components of the reducible 
divisor $(Pfaff)$, cf.~(\ref{Example 1}), (\ref{Example 2}) below. 
In refinement of the criterion $Pfaff(t)=0 \Llra h^0(c_t, \cL_{c_t})\neq 0$, 
one gets for the order with which $Pfaff=\det \iota_1$
(cf.~equ.~~(\ref{extrinsic sequence})) vanishes at $t=t^*$
\beqa
\label{basic inequality}
\mbox{ord}_{t^*}\det\iota_1&\geq &
\dim \ker \iota_1|_{t=t^*}=h^0(c_{t^*}, \cL_{c_{t^*}})
\eeqa
as the number of vanishing eigenvalues of $\iota_1$, i.e.~the algebraic 
multiplicity of the eigenvalue $0$, is $\geq $ its geometric multiplicity.
So, if $h^0(c_t, \cL_{c_t})= m$ generically for $f(t)=0$ 
then (if - as is usually the case and we assume -
$f\neq g^k$ with $k>1$, i.e.~$\mbox{ord}_{t} f =1$) one has $f^m|Pfaff$ (as 
$\mbox{ord}_{t} \det \iota_1=\mbox{ord}_{t}  f\cdot\mbox{mult}_f \det \iota_1$)
and one has for the precise multiplicity 
\beqa
\label{basic estimate}
k'\,:= \, \mbox{mult}_f Pfaff &\geq & 
k_{(f)}\, :=\, h^0(c_t, \cL_{c_t})|_{f(t)=0, \, generic}
\eeqa

As $h^0(c_t, \cL_{c_t})$ is upper semicontinuous we know also 
the following  implication 
\beqa
\label{upper semicont}
\Sigma\subset (f) & \Lra & k_{\Sigma} \; \geq \; k_{(f)}
\eeqa
To infer from (\ref{basic estimate}) and $k_{(f)}$ a multiplicity 
$k'=\mbox{mult}_f\, Pfaff>1$ (i.e.~an interesting {\em lower} bound) 
one would need however an opposite inequality, 
say a {\em lower} bound on $k_{(f)}$ in (\ref{upper semicont}). 
We will get an {\em upper} bound for $k'$ from 
consideration of degrees, cf.~(\ref{consideration of degrees}).

{\em The multiplicity along $\Sigma$}

We are particularly interested in the multiplicity 
$h^0(c_t, \cL_{c_t})$ along the locus $\Sigma$
(which can also have higher codimension) for which 
effective bounds can be derived. The proof in (\ref{cL specialisation}) 
of $\Sigma\subset (Pfaff)$ used a representation of 
$\cL_{c_t}$ for $t\in \Sigma$ as $\cO_{c_t}(D)$ with $D$ effective 
(where of course $\deg D= \deg K_c^{1/2}=g_c-1$). So, by Cliffords theorem,
\beqa
\label{upper bound}
t\in \Sigma \;\;\;\;\; \Longrightarrow \;\;\;\;\;
h^0(c_t, \cL_{c_t})&\leq & \frac{n}{2}\Big(r-\frac{n+1}{2}\Big) +1
\eeqa

On the other hand the specialisations in (\ref{cL specialisation})
allow also to derive a lower bound of $h^0(c_t, \cL_{c_t})$ for $t\in \Sigma$.
In both cases one finds by consideration of the long exact sequence that
the sections of the corresponding bundle on $\E$ in (\ref{cL specialisation})
{\em inject} into the corresponding sections of the bundle restricted to $c_t$.
This gives the following estimates ($m:=r-\frac{n+1}{2}$)
\beqa
\label{lower bound}
t\in \Sigma \;\;\;\;\; \Longrightarrow \;\;\;\;\;
h^0(c_t, \cL_{c_t})&\geq & \left\{ \begin{array}{ll}
\frac{1}{2}\frac{n}{2}(r-\frac{n}{2})+1 & \;\;\;\;\;\;
\mbox{for} \;\;  n\equiv 0 \, (2) \\
\;\;(r-\frac{n+1}{2})+1 & \;\;\;\;\;\; 
\mbox{for} \; \; n\not\equiv 0 \, (2)
\end{array} \right.
\eeqa
where we evaluated $h^0(\E, \cO_{\E}(\frac{n}{2}s+\frac{r-1}{2}F))
=h^0(b, \cO_b(\frac{r-1}{2})\oplus \bigoplus_{i=2}^{n/2}\cO_b(\frac{r-1}{2}-i))
=\frac{r-1}{2}+1+\sum_{i=2}^{n/2}(\frac{r-1}{2}-i+1)=\frac{n}{2}(\frac{r-1}{2}
+1)-(\frac{1}{2}\frac{n}{2}(\frac{n}{2}+1)-1)$ in the first line (note $r>n$);
so
\beqa
\label{even bounds}
\frac{1}{2}\frac{n}{2}(m+\frac{1}{2})&\leq & h^0(c_t, \cL_{c_t})-1 \; \leq \;
\frac{n}{2}m\;\;\;\;\;\;\;\;\;\;\;\;\;\;\;\; 
\mbox{for} \;\;  n\equiv 0 \, (2) \\
\label{odd bounds}
m & \leq & h^0(c_t, \cL_{c_t})-1 \; \leq \; \frac{n}{2}m
\;\;\;\;\;\;\;\;\;\;\;\;\;\;\;\; \mbox{for} \;\;  n\not\equiv 0 \, (2) 
\eeqa

\subsection{\label{The SU(3) case}The case of $SU(3)$ bundles}

\subsubsection{The relation $\; \R=\Sigma$ for $SU(3)$ bundles}

Here one has from the spectral equation 
\beqa
C_rz+B_{r-2}x+A_{r-3}y&=&0
\eeqa
that\footnote{we will not make any notational distinction between the section
$z$ over $\E$ and its restriction to $c_{t}$} 
$z\in H^0(c, \cO_{\E}(3s)|_c)$ has the vanishing divisor 
(where the $u_j$ are the zeroes of $A_{r-3}$)
\beqa
(z)=3s|_c=\sum_{j=1}^{r-3}\, \{3p_0, u_j\}
\eeqa

Now let $t\in \Sigma$ and let 
$(0\neq) \, 
\zeta\in H^0(c_t, \cO_{\E}(3s-\sum_{j=1}^{r-3}\, F_{u_j})|_{c_t})$. 
As the flat bundle of which $\zeta$ is a 
nontrivial section must be trivial $\zeta$ remains everywhere nonvanishing.
Therefore\footnote{recall that a holomorphic/meromorphic section is a 
collection of holom./merom.~functions transforming with suitable 
transition functions 
(and a product of sections is a section of the product bundle)}
$z/\zeta$ is not only an element of 
$\Gamma_{mero}(c, \cO_{\E}(\sum_{j=1}^{r-3}\, F_{u_j})|_{c_t})$
but one has even 
\beqa
\frac{z}{\zeta}&\in & 
H^0\Big(c, \cO_{\E}\Big(\sum_{j=1}^{r-3}\, F_{u_j}\Big)\Big|_{c_t}\Big)
\eeqa
Now tensoring the short exact sequence $0\lra \cO_{\E}(-c_t)\lra \cO_{\E}\lra
\cO_{c_t}\lra 0$ with $\cO_{\E}((r-3)F)$ and taking the long exact
cohomology sequence tells one that actually
\beqa
H^0\Big(c, \cO_{\E}\Big(\sum_{j=1}^{r-3}\, F_{u_j}\Big)\Big|_{c_t}\Big)&=&
\pi^* H^0\Big(b, \cO_b(r-3)\Big)
\eeqa
because $H^1(\E, \cO_{\E}(-3s-3F))=0$, cf.~equ.'s (C.22) and (C.17)
of [\ref{extrinsic}]. Therefore one gets that
$z/\zeta|_{c_t}=\pi^* P_{r-3}|_{c_t}$ and thus
(with the $u_k'$ denoting the $r-3$ zeroes of $P_{r-3}$)
\beqa
\label{divisor equality}
\sum_{k=1}^{r-3}F_{u'_k}|_{c_t}&=&(\pi^*P_{r-3})|_{c_t}
\; = \; \Big(\frac{z}{\zeta}\Big)\Big|_{c_t}\; = \; 3s|_{c_t}
\eeqa
such that, whereas (\ref{Sigma relation}) and (\ref{curly R relation}) 
gave $\R\subset \Sigma$, one gets  here also that $\Sigma\subset \R$
(as $\R$ is, 
cf.~(\ref{curly R relation}), the locus of $t\in \M_{\E}(c)$ where $3s|_{c_t}$ 
equals the sum of the $r-3$ fibers $F_{u_j}|_{c_t}$ of $\pi_{c_t}:c_t\lra b$ 
over the zeroes of $A_{r-3}$; now the fibers $F_{u'_k}|_{c_t}$
in (\ref{divisor equality}) must be necessarily these
as they must contain the $\{p_0, u_j\}$). So 
for $SU(3)$ bundles we find finally that 
\beqa
\R=\Sigma
\eeqa
In general, for $SU(3)$ bundles one has 
for the multiplicity 
$k_{\Sigma}=h^0(c_t, \cL_{c_t})$ along $\Sigma$
\beqa
\label{SU(3) bounds for kSigma}
r-1\leq k_{\Sigma} &\leq &\frac{3}{2}r-2
\eeqa

Let us recall the locus
$\R$ (we switch between both notations:
$a_2=B_{r-2},\, a_3=A_{r-3}$)
\beqa
\R&=&\bigcap_{j=1}^{r-3}(R_2^{(j)})\; =
\; \bigcap_{j=1}^{r-3}(Res(a_2, a_3^{(j)}))
\eeqa
of codim $r-3$,
i.e.~the locus where $A_{r-3}|B_{r-2}$ or
\beqa
\label{divisibility locus}
\Sigma\, = \, \R&=&\Big\{t\in \M_{\E}(c)\Big| \, 
\exists \, L_1\; s.t.\;\; B_{r-2}=L_1A_{r-3}\Big\}
\eeqa
Here codim $\Sigma= r-3$, so $= 1,2$ for $r=4,5$, 
resp.~(cf.~(\ref{Example 1}), (\ref{Example 2}) 
and Ex.'s 1, 2 [\ref{extrinsic}]). 

Here the locus $R$, cf.~app.~\ref{aux locus}, 
is a $\leq $ codim $1$ subspace (cf.~sect.~3.2.2 of [\ref{extrinsic}])
where $(3s-F)|_{c_t}\sim$ effective holds;
the latter by R-R holds always for 
$r\geq 6$ and for $r=5$ (or $r=4, \la>\frac{3}{2}$, cf.~after (\ref{r})) 
poses a condition. 

\subsubsection{$SU(3)$ bundles with $\la = 3/2$ and the special locus $(f)$}

Let us describe one important special case in a more detailed manner.
For $\la = 3/2$ 
one has $l(-F)=\cO_{\E}(6s-(r-4)F)$ and we took, cf.~[\ref{extrinsic}],
\beqa
\label{lbar bundle}
\bar{l}(-F)=\left\{ \begin{array}{ll}
\cO_{\E}(3s-\frac{r-4}{2}F)& \;\;\;\;\;\;
\mbox{for} \;\;  r\; \mbox{even} \\
\cO_{\E}(3s-\frac{r-3}{2}F)& \;\;\;\;\;\; 
\mbox{for} \; \; r \; \mbox{odd}
\end{array} \right.
\eeqa
We consider $\iota_1: H^1(\E, \widetilde{\cL}(-c))\ra
H^1(\E, \widetilde{\cL})$ with $\widetilde{\cL}=l(-F)$ 
and the corresponding map $\bar{\iota}_1$ for $\bar{l}(-F)$.
We got $f|Pfaff$ for 
\beqa
Pfaff&=&\det \iota_1\\
f    &=&\det \bar{\iota}_1
\eeqa
For $r$ even one has $f\equiv 0$ as $h^0(c, \bar{l}(-F)|_c)=-3(\frac{r}{2}-1)
+h^0(c, \cO_{\E}(3(\frac{r}{2}-1)F)|_c)\geq 1$, such that also $Pfaff\equiv 0$
as $h^0(c, l(-F)|_c)\geq 1$ by $l(-F)|_c=\bar{l}(-F)|_c^{\ox 2}$.
For $r$ odd the matrix representation (G.12) of 
[\ref{extrinsic}] for $\bar{\iota}_1$ shows (via linear dependence of the rows)
\beqa
(f)&=&\Big\{t\in \M_{\E}(c)\Big| \, \exists \,
\tilde{C}_k, \tilde{B}_{k+2}, 
\tilde{A}_{k+3}, \mbox{{\normalsize $k=\frac{r-7}{2}$}},  s.t.\;\;
\tilde{C}C_r+\tilde{B}B_{r-2}+\tilde{A}A_{r-3}=0\Big\}  \;\;\;\;\;\;\;\;
\eeqa
(where $\tilde{C}=0, (f)=R$ for $r=5$). 
So one gets from (\ref{divisibility locus})
(cf.~also (\ref{n=3 la=3/2}), (\ref{Example 1}) below)
\beqa
\Sigma \, = \, \R&\subset &(f)\; \subset \; (Pfaff)
\eeqa
(by using $\tilde{C}=0, \tilde{B}\equiv 1, \tilde{A}=-L_1$).
If (for $r$ odd) now $t\in (f)$, 
i.e.~if a nonzero element 
$\rho \in H^0(c, \bar{\cL}_t)$ exists, 
one gets linearly independent elements 
$\rho^2 u, \rho^2 v\in H^0(c, \cL_t)$, so
\beqa
\label{kf estimate}
k_{(f)}&\geq & 2
\eeqa

\subsubsection{\label{more specialised information}
More specialised information for certain $SU(3)$ bundles}

In [\ref{extrinsic}] we looked, following [\ref{BDO}], for components $k'(f)$,
where $k'=\mbox{mult}_f\, Pfaff$,
of the reducible vanishing divisor $(Pfaff)=(\det \iota_1)$ of the Pfaffian. 
The factor $f$ (respectively $Res(a_2, a_3)$ in Example 2) arose as a
$\det \bar{\iota}_1$ for a related line bundle $\bar{l}(-F)$. 
We get
\beqa
\label{consideration of degrees}
k'\, h^1\Big(\E, \bar{l}(-F)\Big)&\leq &h^1\Big(\E, l(-F)\Big)
\eeqa
as an upper bound on $k'$ from consideration of degrees 
and in [\ref{extrinsic}] we
got (for $\chi=1$) the following results\footnote{besides an $SU(4)$ case
of $r=9, \la=1$ with $\cL=\cO_{\E}(6s-F)|_c$ and 
$(Pfaff_{(20)})=(f_{(9)})+(g_{(11)})$
(where $\bar{l}(-F)=\cO_{\E}(4s-F)$)}
(subscripts in brackets indicate the polynomial degree):

\un{(A) $r\geq 5$ odd, $\la=3/2$} with $\cL=\cO_{\E}(6s-(r-4)F)|_c$ and
(with $f$ from (\ref{lbar bundle}))
\beqa
\label{n=3 la=3/2}
(Pfaff_{(6r-10)})&=&k'(f_{(\frac{3r-5}{2})})+(g)
\eeqa
Here $4\geq k'\geq k_{(f)}\geq 2$
from consideration of degrees, (\ref{basic estimate})
and (\ref{kf estimate})
(and $k_{\Sigma}\geq k_{(f)}$ with $k_{\Sigma}\geq 4$
by $\Sigma \subset (f)$, (\ref{upper semicont}) 
and (\ref{SU(3) bounds for kSigma})), so \fbox{predicting $k'\geq 2 $}.

\un{(B) $r=5, \la = 5/2$} with $\cL=\cO_{\E}(9s-3F)|_c$ and
(with $f=\det \bar{\iota}_1=Res(B_3, A_2)$ 
where $\bar{l}(-F)=\cO_{\E}(3s-F)$
and $k'\leq 12$ from degrees while $k_{\Sigma}=4$ or $5$)
\beqa
(Pfaff_{(62)})&=&k'(f_{(5)})+(g)
\eeqa

We considered, with the notation $\Sigma=(f_{\Lambda})$,
in greater detail the following cases [\ref{extrinsic}]:

\un{Example 1: $r=4, \la = 5/2$} with $\cL=\cO_{\E}(9s-F)|_c$ and
(with $R$ the zero-divisor of the polynomial $R_2=Res(a_3, a_2)$
which is of degree $3$) (cf.~also app.~\ref{aux locus})
\beqa
\label{Example 1}
(Pfaff_{(44)})&=&k'\, R_{(3)}+(Q_{(44-3k')})
\eeqa 
Here $(R_2)=R=\R=\Sigma$ and 
$\mbox{codim} \, \R=\mbox{codim} \, \Sigma=1$ from (\ref{Sigma remark 3}),
so $\Sigma$ is a component of $(Pfaff)$.
As here $\Lambda=\cO_{\E}(3s-F)$ one gets 
that $\cL\cong \cO_{\E}(2F)|_c$ along $\Sigma$
such that $3\leq k_{\Sigma}\leq 4$, \fbox{predicting $k'\geq 3$}
by (\ref{basic estimate}) (with $\Sigma=(R_2)$) 
in accord with $k'=11$, cf.~[\ref{BDO}].

\un{Example 2: $r=5, \la = 3/2$} (a special case of (A) above)
with $\cL=\cO_{\E}(6s-F)|_c$ and 
\beqa
\label{Example 2}
(Pfaff_{(20)})&=&k' R_{(5)}+(g_{(20-5k')})
\eeqa
where $R=(R_2)=(Res(a_2, a_3))$ such that $(3s-F)|_{c_t}\sim$ 
effective along $R$ (cf.~app.~\ref{aux locus}); 
for $r=5$ this is enough for $(6s-(r-4)F)|_{c_t}\sim$ effective, 
i.e.~$f|Pfaff$ or $R=(f)\subset (Pfaff)$;
but one has not yet $(3s-(r-3)F)|_{c_t}\sim$ effective for 
$\Lambda=\cO_{\E}(3s-(r-3)F)$, which needs for $r=5$ a second condition
to be posed (so $\mbox{codim}\, \Sigma=2$).
Here one has $4\geq k'\geq k_R\geq 2$ 
from consideration of degrees, (\ref{basic estimate})
and (\ref{kf estimate}) (and $k_{\Sigma} \geq k_R$ 
with $4\leq k_{\Sigma} \leq 5$
by $\Sigma=\R\subset R$, (\ref{upper semicont})
and (\ref{SU(3) bounds for kSigma})),
so \fbox{predicting $k'\geq 2 $} and concretely $k'=4$, cf.~[\ref{BDO}].

\section{\label{Bundles over b}The moduli space over the instanton curve}

\resetcounter

The contribution criterion (\ref{main contribution criterion}), 
and even the precise prefactor, depends only on the
restriction $V|_b$ of $V$ to the instanton curve $b$,
i.e.~the Pfaffian factors through the map
\beqa
\M_X\, (V)&\stackrel{\rho_b}{\lra} & \M_b\, (V_b)
\eeqa
(defined by $V\ra V|_b$) to the moduli space 
of $SU(n)$ bundles $V_b$ over $b$. By Grothendieck's decomposition theorem
one has
(the array of the $k_i$ is called the splitting type) 
\beqa
\label{Grothendieck decomposition}
V_b&=&\bigoplus_{i=1}^n\, \cO_b(k_i)
\eeqa
This concerns $Gl(n, {\bf C})$ or, equivalently, $U(n)$ bundles.
For $SU(n)$ bundles 
we get the further condition $\int_b c_1(V|_b)=\sum_{i=1}^n\, k_i\, = \, 0$,
giving the space of sets of $n-1$ integers as moduli space:
considering the lattice $\Gamma_{SU(n)}$
\beqa
0\; \lra \; \Gamma_{SU(n)}\; \lra \; {\bf Z}^n \;
\stackrel{\sum}{\lra} \; {\bf Z}\; \lra \; 0
\eeqa
with the natural symmetry action of $W_{SU(n)}$, the symmetric group in 
$n-1$ elements, gives the moduli space (the complement of the divisor 
$(Pfaff)$ in $\M_X(V)$ is $\rho_b^{-1}\big( (0, \dots , 0) \big)$)
\beqa
\M_b(V_b)& = & \Gamma_{SU(n)}/W_{SU(n)}
\eeqa

A world-sheet instanton supported on $b$
contributes according to the criterion [\ref{Wittenhet}] 
\beqa
\label{main criterion}
W_b\neq 0 & \Llra & h^0\Big(b, V|_b(-1)\Big)=0
\eeqa
with $V|_b(-1):=V|_b\ox \cO_b(-1)$
and $h:=h^0\big(b, V|_b(-1)\big) =  \sum_{k_i> 0} \, k_i$.
So $h=0\Leftrightarrow  k_i=0$, for all $i$, i.e.~$b$ contributes precisely 
if $V|_b$ is 
trivial: $W_b\neq 0 \Longleftrightarrow  V|_b = \bigoplus_{1}^n\, \cO_b$;
and $h=h^0\Big(b, \pi_{c*}l|_c\otimes \cO_b(-1)\Big)= h^0(c, \cL)$.
One has\footnote{note 
$h^0\big( b, V|_b(-1) \big)- h^1\big( b, V|_b(-1) \big)
=\int_b c_1\big(V|_b(-1)\big)+\frac{c_1(b)}{2}=\int_b c_1(V)|_b=0$
such that
$0=\deg V|_b = \chi(b, V|_b(-1)) = \chi(c, l(-F)|_c)= 
\deg l(-F)|_c -  \deg K_c^{1/2}$
(even $h^i(b, V|_b(-1))=h^i(c, l(-F)|_c)$).} 
$\cL =K_c^{1/2}\ox \cF$,
with the flat $\cF=\Lambda|_c^{\la}$ for $n$ even; for $n$ odd 
the decomposition depends on a spin choice with its $Pic_0^2$ ambiguity.
(For $SU(2)$ bundles 
$V|_b=\cO_b(h)\oplus \cO_b(-h)$ with 
$h:=h^0(c, \cL)=h^0(b, V|_b(-1))$.\footnote{The jumping phenomenon 
arises if a one-parameter family 
of spectral curves $c'\in |c|$, with a corresponding family of line bundles 
$\widetilde{\cL}|_{c'}\in Pic_{g-1}({c'})$, meets 
$\Theta_{c'}$ at $t=0$, say at a generic smooth point: 
then $V|_b$ is $\cO_b\oplus \cO_b$
for $t\neq 0$ and $\cO_b(h)\oplus \cO_b(-h)$ at $t=0$ where $h=1$.})

By (\ref{Sigma remark 1}),(\ref{Sigma remark 2}) one gets, in $\M_{\E}(c)$,
on {\em the complement of} $(Pfaff)$ (contained in the complement of $\Sigma$)
that $k_i=-k_i$ for all $i$, and on $\Sigma_{2\la}(\supset\Sigma)$ 
that $\{ k_i \}=\{ - k_i \}$ as sets, i.e.~there one has
\beqa
\label{self-dual decompositions}
V|_b=\left\{ \begin{array}{ll}
\bigoplus_{i=1}^{n/2} \cO_b(k_i) \oplus \cO_b(-k_i)  \;\;\;\;\;\;\;\;
& \mbox{for} \;\; n\equiv 0 \; (\, mod \, 2) \\
\cO_b \oplus \bigoplus_{i=1}^{(n-1)/2} \cO_b(k_i) \oplus \cO_b(-k_i) 
\;\;\;\;\;\;\;\;
& \mbox{for} \;\; n\not\equiv 0 \; (\, mod \, 2)
\end{array} \right.
\eeqa
$(Pfaff)\subset\M_{\E}(c)$, the set where a $k_i\neq 0$ exists, contains
$\Sigma$ (where $\{ k_i \}=\{ - k_i \}$).

\section{\label{intrinsic}Intrinsic derivation of $Pfaff$ 
on the instanton curve}

\resetcounter

In [\ref{extrinsic}] we stated the following necessary criterion for 
contribution to the superpotential
\beqa
\label{neces cond}
W_b \; \neq \; 0 \; \Longrightarrow \; \beta \; < \; 0
\eeqa
(cf.~equ.~(4.27) of [\ref{extrinsic}]);
recall the notation $\widetilde{\cL}=l(-F) 
= \cO_{\E}(n(\la + \frac{1}{2})s+\beta F)=:\cO_{\E}(\widetilde{D})$.
This was done via an extrinsic detour over the sheafs 
$\widetilde{\cL}\otimes \cO_{\E}(-c)$ and $\widetilde{\cL}$ on
$\E$. Actually there is also a more direct, intrinsic argument, 
using directly the sheaf $\cL=\widetilde{\cL}|_c$ on $c$.

Recall that $h^0(c, \cO_c(D))-1$ is the (projective) dimension 
of the associated linear system $|D|$
(of effective divisors linear equivalent to $D$) of a divisor $D$ on $c$ 
(with $\cO_c(D)$ here as the associated line bundle on $c$). So one recovers 
immediately (\ref{neces cond}) by (cf.~(4.35) of [\ref{extrinsic}])
\beqa
\label{corol}
h^0(c, \cO_c(D))=0 & \Llra & D \; \not\sim \; \mbox{effective}
\; \Lra \; D \; \neq \; \mbox{effective}
\eeqa
as $W_b\neq 0 $ means according to the precise criterion (\ref{main criterion})
just $h^0(c, \cL)=0$. 

One has the representation 
$\cL =  K_c^{1/2}\otimes \cF \; = \; \cO_c(D)=\cO_{\E}(\widetilde{D})|_c$
of the relevant line bundle as being 
composed\footnote{\label{notation adjusted for n odd}this is for $n$ even; 
for $n$ odd a corresponding decomposition depends on a spin choice with its 
$Pic_0^2$ ambiguity; the assertion, and what follows, 
then has to be somewhat adjusted in notation, 
cf.~discussion after (\ref{l(-F) on E}),(\ref{the decomposition}) 
and footn.~\ref{integralities}, as the factors of $\cL$ then do not arise
by restriction of line bundles on $\E$} 
of the spin bundle of $c$ and a flat bundle 
with $d:=\deg D = g-1$ and $g:=g_c$ and $\cF_t=\cO_{c_t}(G|_{c_t})$ 
with $G=\lambda(ns-(r-n\chi)F)$ for a $V_t$ with $t\in \M_{\E}(c)$. Recall the 
map\footnote{for $\omega_{\al}$ a basis of holomorphic differentials
normalized by $\int_{a_{\al}}\omega_{\beta}=\delta_{\al \beta}$ with respect to
a canonical symplectic basis $(a_1, \dots , a_g, b_1, \dots, b_g )$ 
of $H^1(c, {\bf Z})$, i.e. $a_{\al}\cdot a_{\beta}=b_{\al}\cdot b_{\be}=0, \,
a_{\al}\cdot b_{\beta}=\delta_{\al \beta}$; recall $Jac(c)={\bf C^g}/\Lambda$
with the lattice $\Lambda$ generated by the $a\,$- and $b\,$- periods of
the $\omega_{\al}$; $\mu$ is still well-defined on divisor classes, 
cf.~app.~\ref{jacobian} where 
the Riemann theta-function $\theta$ and its divisor $\Theta$ are recalled}
$D\ra \mu(D)=\sum_i \mu(p_i)=
(\sum_i \int_{p_0}^{p_i} \omega_{\al})_{\al =1, \dots ,g}$ 
(the point $p_0$ chosen fixed) first 
from effective divisors $D=\sum_{i=1}^d p_i$ (then prolonged to all divisors)
to the Jacobian 
\beqa
\label{mu map}
\mu: \; Sym^{g-1}c \; \lra \; Jac(c)
\eeqa
of image $W_{g-1}=\Theta - \kappa$ with $-\kappa = \mu(\frac{1}{2}K_c)$
by a theorem of Riemann
(with $W_{d}:=\mu(Sym^{d} c)$ 
and $\frac{1}{2}K_c$ understood as divisor class); 
enhancing $\mu$ to a map from all divisors (cf.~app.~\ref{Jacobi})
one finds that the preimage
of $W_{g-1}$ are those divisors of degree $g-1$ which are $\sim $ effective.
Then one has, as $W_b\neq 0 \Llra \frac{1}{2}K_c + G|_c \not\sim $ effective 
by (\ref{main criterion}) and (\ref{corol}),
\beqa
\label{theta Pfaff relation}
Pfaff\, ( \delbar_{V_t|_b(-1)})=0& \Llra & 
\theta \Big( \mu ( G|_{c_t}), \Omega_t \Big) =0 
\eeqa
(cf.~(\ref{Omega t definition})).
The fact that $\theta$ is an even function 
$\theta(-z, \Omega)=\theta(z, \Omega)$
corresponds to the fact that $\cL_{-\la}$ contributes exactly if
$\cL_{\la}$ does as $h^0(c, \cL_{\la})=h^0(c, \cL_{-\la})$, cf.~equ.~(E.3)
in [\ref{extrinsic}].

\subsection{Relation of the extrinsic and intrinsic considerations}

We consider the bundle $\underline{Jac}$ of 
Jacobians over the moduli space $\M_g$, cf.~app.~\ref{Jacobi}. 
In each fibre $Jac(c_{\Omega})$ one 
has the respective theta divisor $\Theta_{\Omega}$ of $\theta(\cdot, \Omega)$. 
These divisors fit together to a divisor $\Theta$
in the total space $\underline{Jac}$.
In our concrete situation the extrinsic moduli space is 
embedded (for $\Pi$ an injective immersion) in the base, via a map $\Pi$
\beqa
\label{Omega t definition}
{\bf P}H^0(\E, \cO_{\E}(c))\; \ni \; t & \stackrel{\Pi}{\lra} & \Pi(t)\; = \; 
\Omega_t\; \in \; \M_g
\eeqa
with image $I$, say. Here $t$ stands for the specific, concrete curve 
$c_t\in |c|$ inside $\E$ (which is parametrized by $t$ in the moduli space)
given by the zero set $(s_t)$ of a section $s_t$ in $H^0(\E, \cO_{\E}(c))$.
Now let us enhance the perspective to the respective full fibrations
\beqa
\label{one-dim situation with abstract fibration}
\M_{\E}(V_{\E})\;\;\;\;\;\;\; \lra \;\;
\underline{Jac}  \;\;\;\;\;\;\;\;\;\nonumber\\
\;\;\;\; \da \; Jac( \cdot ) \;\;\;\;\;\;\;\;\;\; p \da \; Jac(\cdot)\\
{\bf P}H^0\Big(\E, \cO_{\E}(c)\Big)\;\;\; \stackrel{\Pi}{\lra} \;\; \M_g
\;\;\;\;\;\;\;\;\;\nonumber
\eeqa
Using the additional information provided by the line bundle 
$\widetilde{\cL}|_{c_t}$ over $c_t$ one gets ($\la$ held fixed) 
a section\footnote{Note that, contrary to a situation
where one has a family of linear equivalent divisors on one fixed curve which 
are mapped to just one point in a Jacobian, here one has one divisor $\D$ 
(in the ambient space $\E$, with $\widetilde{\cL}=\cO_{\E}(\D)$) 
which restricts to divisors on the respective members of 
a family $|c|$ of linear equivalent curves; the ensuing family of divisors
is then mapped to Jacobian fibre points over the respective base points in 
$I\subset \M_g$ (themselves corresponding to the different members in $|c|$).}
$\cS_{\la}$ over $I=im\, \Pi \subset \M_g$ 
of the $\underline{Jac}$-fibration from the section 
one started with in the fibered space $\M_{\E}(V_{\E})$
(which itself came from the original $V$ over $X$)
\beqa
\label{transfer 2}
\Gamma\Big( {\bf P}H^0\big( \E, \cO_{\E}(c)\big), \M_{\E}(V|_{\E}) \Big)
\; \ni \; S_{\la} \;\; & \lra & \;\; \cS_{\la}\; \in \;
\Gamma\Big( I, \underline{Jac}\Big)\;\;\;\;\;\;
\eeqa
(a global spin choice made); so this continues the map of fibrations in 
(\ref{transfer 1}). In other words, one has a (choice-dependent) 
mapping\footnote{Here we give the fibre coordinates first as is 
customary in connection with the theta function $\theta(z, \Omega)$ which 
itself is defined on the set of $(z, \Omega_t)\in \underline{Jac}$. 
Furthermore, by abuse of language, we use the same symbol for a section 
and the function giving the fibre coordinate in dependence of the base point.} 
from $t\in {\bf P}H^0(\E, \cO_{\E}(c))$ to corresponding pairs
$(z(\Omega_t), \Omega_t)$ where $\Omega_t=\Pi(t)\in I\subset \M_g$ and 
$z(\Omega_t)=\mu(G_{c_t})\in Jac(c_{t})$ ($n$ even, say)
\beqa
\label{map to Jac}
\M_{\E}(V_{\E})\; &\ni &\Big( S_{\la}(t)\, ,\,  t\Big)\;\;\; =\;\;
\Big( \cL_{c_t}, t\Big) \nonumber\\
&\lra & \Big( \cS_{\la}(\Omega_t), \Omega_t \Big) =
\Big( \mu(G_{c_t}), \Omega_t\Big) =
\Big([z_t], \Omega_t\Big)\; \in \; \underline{Jac}
\eeqa
Considering just the part $\underline{Jac}|_I$ and the corresponding restricted
global theta divisor $\Theta|_I$, 
the question is where (i.e.~at which points of $I$) 
the section $\cS_{\la}$ meets that divisor
\beqa
p(\cS_{\la}\cap \Theta|_I)
\eeqa
One expects therefore an identification of the following divisors
(with $z(\Omega)$ as described)
\beqa
\label{comparison of hypersurfaces}
{\cal H}\; = \; 
\Big\{t\, \Big|\, \det \; \iota_1(t) =0 \Big\} 
\subset {\bf P}H^0\big(\E, \cO_{\E}(c)\big)
& \stackrel{\Pi}{\lra} & \Big\{\Omega\in I\, \Big| \, 
\theta\big( z(\Omega), \Omega\big)=0\Big\}\subset I\subset \M_g
\;\;\;\;\;\;\;\;\;
\eeqa

The fibre has dimension $g_c$. 
But the whole space of vertical moduli in each respective fibre $Jac(c_t)$,
i.e.~of all flat line bundles on $c_t$, 
is not at our disposal as arguments for $Pfaff$; rather,
the available variation stems just from a movement in the base $I$
(which comes from motions in $|c|$, i.e.~motions of $c$ in $\E$, that is
moduli of the restricted bundle); in each fibre only the section point of 
$\cS_{\lambda}$ occurs as argument of $Pfaff$. Therefore, whereas an 
a priori {\em intrinsic} 
investigation would be concerned with a divisor condition
in the full {\em vertical} (Jacobian) direction, {\em extrinsically} one 
investigates the condition inside $|c|$, respectively its image $I$ 
in the base $\M_g$, i.e. along a {\em horizontal} (base) direction.

\noindent
One expects by (\ref{comparison of hypersurfaces}) 
even the identification of a transcendental function as a polynomial 
\beqa
\label{function equality}
\det \; \iota_1\, (t) \; & = & \; 
\theta\Big( \cS_{\la}\big(\Pi(t)\big), \Pi(t)\Big)
\eeqa
with the $\theta$-function
(here actually $\cS_{\la}\big(\Pi(t)\big)$ gives only 
$[z(\Omega_t)]\in Jac(c_t)$) and not $z(\Omega_t)$)
\beqa
\theta\Big( \cS_{\la}\big(\Pi(t)\big), \Pi(t)\Big)\; = \; 
\sum_{l\in {\bf Z^g}}\, 
\exp\Big\{2\pi i \Big( \frac{1}{2}\Big\langle l, \Pi(t)l\Big\rangle
+ \Big\langle l, \cS_{\la}\big(\Pi(t)\big)\Big\rangle \Big)\Big\}
\eeqa
where we recall that\footnote{with $(a_i, b_i)$, $i=1, \dots, g$, a canonical
symplectic basis of $H_1(c_t, {\bf Z})$, 
'constant' under varying $t$} 
(for explicitness we describe the case $n$ even 
where $\cF_{c_t}=\cO_{c_t}(G|_{c_t})$, 
so that we do not discuss integrality-issues and $2^{2g}$-ambiguity, 
cf.~footn.~\ref{notation adjusted for n odd})
\beqa
\Pi(t)&=&\Omega_t\in \cH_g\;\;\;\;\;\;\;\; \;\;\;\;\;\;\;\;\;\;\;
\mbox{with}\; \;\;\Omega_t\, = \, \Big(\int_{b_{\beta}}\omega_{\alpha}\Big)\\
\cS_{\la}\Big(\Pi(t)\Big)&=&[z_t]\in {\bf C^g}/\Lambda
\;\;\;\;\;\;\;\;   \mbox{with}\; \;\;
[z_t]\, = \, \mu(G|_{c_t})
=\Big(\sum_{i=1}^{\la n(r-n)} \int_{q_i}^{p_i} 
\omega_{\alpha}\Big)_{\alpha=1, \dots, g}
\;\;\;\;\;\;\;\;
\eeqa
Here the $\omega_{\alpha}$, $\alpha = 1, \dots, g$ are 
a basis of $H^{1,0}(c_t)$ (cf.~remark after (\ref{genus})
how they arise concretely by taking Poincare residues),
normalized by $\int_{a_{\beta}}\omega_{\alpha}=\delta_{\alpha\beta}$.
Furthermore one has
(where $a_n(u_i)=0$, cf.~app.~\ref{coordinates})
\beqa
G|_{c_t}&=&\la n s|_{c_t}- \la (r-n)F|_{c_t}\; = \; 
\sum_{i=1}^{\la n (r-n)}\Big(p_i-q_i\Big)\;\;\;\; \mbox{where}\;
\left\{ \begin{array}{ll}
s|_c = \sum_{i=1}^{r-n} \{p_0, u_i\}\\
F_{u_*}|_c = \sum_{i=1}^n \{ q_i, u_*\}
\end{array} \right. \;\;\;\;\;\;
\eeqa
and where $z_t$ is only well-defined up to the periods, 
i.e.~as $[z_t]\in Jac(c_t)\cong {\bf C^g}/\Lambda$.
Here $z_t$ arises from an object $G$ on $\E$ which remains constant
for a varying $t\in \M_{\E}(c)$.

As $\Sigma$ denotes the subset of $\M_{\E}(c)$ 
where $(ns-(r-n)F)|_{c_t}\sim 0$ one has also (cf.~(\ref{Sigma remark 2}))
\beqa
\Sigma\; \subset \;\{Pfaff(\cdot)=0\} &\lra &
\cS_{\la}(\Pi(\Sigma))\subset\Theta
\eeqa

\subsection{Some further remarks about multiplicities}

We conclude these comparisons by continuing the considerations 
in sect.~\ref{multiplicities}:  we now want to compare the perspectives 
which the extrinsic and the intrinsic approach offer, respectively, 
on the issue of multiplicity.

\subsubsection{Extrinsic point of view}

In the extrinsic consideration one has a moduli space 
$\M_{\E}(c)=|c|={\bf P}H^0(\E, \cO_{\E}(c))$ and 
a line bundle $\widetilde{\cL}|_{c_t}$ 
over each of the respective curves $c_t\in |c|$.
Inside $|c|$ one has a hypersurface 
${\cal H}=\{\det \, \iota_1 = 0\}$ of degree $h^1(\E, \widetilde{\cL})$.
The matrix for the map 
\beqa
\iota_1: H^1\Big(\E, \widetilde{\cL}\otimes \cO_{\E}(-c)\Big)
&\lra &H^1\Big(\E, \widetilde{\cL}\Big)
\eeqa
arises from multiplication with the defining polynomial 
$s_t$ of the curve $c_t=\{s_t=0\}\in |c|$. Therefore the moduli, which are
essentially given by the coefficients of the monomials in $s$,
enter the matrix-elements linearly. 
So $\det \, \iota_1=0$ is one equation (of determinantal form) 
of degree $h^1(\E, \widetilde{\cL})$ in the moduli. That is, in the extrinsic 
approach of equ.~(\ref{extrinsic sequence}), the Pfaffian is identifed 
as a polynomial in the moduli which has the form of a determinant. From the 
determinantal nature of $Pfaff$, with respect to {\em horizontal} variations,
i.e.~variations of the specific curve $c_t$ (which means varying the bundle
$V_T$ over $b$, where $T\in \M_X(V)$ with $T\lra t \lra z_t$), one finds then 
the following. One has the number $h^0(b, S_-(b)\ox V|_b)$ of occuring 
zero-modes, so one gets the refinement of the contribution criterion 
(\ref{main criterion}) that $Pfaff$ vanishes (at least) to order 
$h^0(b, S_-(b)\ox V_T|_b)$, that is one has 
\beqa
ord_t\, Pfaff &\geq & h^0(c_t, \cL_{c_t})
\eeqa
(cf.~equ.~(\ref{basic inequality})).
We wish to emphasize that here the source domain of values for $Pfaff$
is a horizontal one, in the sense described. This is typical for the
extrinsic algebraic approach. In the intrinsic transcendental approach,
one considers the general theta function $\theta(z, \Omega_c)$
of two variables and is usually concerned with its variation
in the first entry, the second one held fixed. The latter, describing the 
isomorphism type of the concrete curve $c$, corresponds to
the mentioned horizontal variations of the concrete curve $c_t\in \M_{\E}(c)$
in the extrinsic approach. By contrast the variable $z$ corresponds under 
the Abel/Jacobi-map $\mu$ to specific divisor classes, or equivalently
line bundles, on a fixed curve (this is what we call here the intrinsic
aspect of the transcendental approach). Despite this difference it is 
interesting to see nevertheless a corresponding role of $h^0(c_t, \cL_{c_t})$
for $\theta(\cdot , \Omega_c)$ in the intrinsic approach.

\subsubsection{Intrinsic point of view}

If $z=\mu(G_{c_t})\in Jac(c)$ corresponds to $t\in\M_{\E}(V|_{\E})$ 
we saw in (\ref{theta Pfaff relation}) 
that $\theta(z)\; = \; 0 \Leftrightarrow  Pfaff(t)\; =\; 0$
as Riemann's theorem $W_{g-1}=\mu(K_c^{1/2})+ \Theta_c$
shows that just then an {\em effective} divisor class
of degree $g-1$ belongs to the line bundle $\cL= K_c^{1/2}\ox \cF_z$
(such that $S_-(c)\ox \cF$ has non-trivial sections 
and $b$ can not contribute to the superpotential).

Actually Riemann did not stop at the relation described so far
but elucidated (in terms of the associated linears systems on the curve)
the finer structure of the analytic subvariety 
given by the theta-divisor, especially its stratification by singular points. 

We saw that for $\mu(G|_c)\in \Theta$ the divisor class 
$\frac{1}{2}K_c+G|_c$ related to $\cL$ was effective, so 
\beqa
z=\mu(\cL)-\mu(K_c^{1/2})\; \in \; \Theta & \Lra & h^0(c, \cL)\; >\; 0
\eeqa
Now recall first the characterisation of the regular locus
\beqa
\mu(D_d)\; \in \; W_d \;\; \mbox{is a smooth point} & \Llra &
h^0(c, D_d)\; =\; 1
\eeqa
This gives in our case the following special case of 
Riemann's singularity theorem
\beqa
z=\mu(\cL)-\mu(K_c^{1/2})\; \in \; \Theta_{reg}& \Llra &
h^0(c, \cL)\; =\; 1
\eeqa
Riemann's theorem asserts 
(where codim  $\Theta_{sing}= 3$ if $c$ is not hyperelliptic; then it is $2$)
\beqa
\label{Riemann's singularity theorem}
\mbox{mult}_z\; \Theta & = & h^0(c, \cL)
\eeqa
Here one considers the $\theta(z, \Omega)$-function 
for fixed argument $\Omega$, i.e.~one considers in the Jacobian fibration
(\ref{Jacobian fibration}) over $\M_g$ just the {\em vertical} direction.

{}From $\theta(z) =  0  \Longleftrightarrow  Pfaff(t) = 0$
above one gets that $Pfaff(t)$ and $\theta(z_t, \Omega_t)$ 
are (up to a constant) powers of each other. We are interested 
in the case where this is a linear relation such that one
actually has 
\beqa
\label{Pfaff theta relation}
Pfaff(t) & = & const. \cdot \theta(z_t, \Omega_t)
\eeqa
The theta divisor, when considered {\em vertically}, 
is described {\em locally} by a determinantal expression of rank
$h^0(c, \widetilde{\cL}|_{c_t})$ at $z(t)$:
for this recall that, by a theorem of Kempf (cf.~(\ref{Kempf})),
in our case of $d=g-1$ for $\cL=K_c^{1/2}\ox \cF_t$,
the codimension one analytic subvariety $\Theta_{-\kappa} = W_{g-1}$ 
can be described near $z=\mu(D)$ by an equation $\det \, f_{ij} =0$
where $f_{ij}$ is an $h\x h$ matrix of functions holomorphic at $z$
where $h=h^0(c, \cO_c(D))$ with specific linear terms.
This refines (\ref{Riemann's singularity theorem})
as thereby (when considered {\em vertically}, i.e.~for fixed $\Omega$)
the divisor $\Theta_{(-\kappa)} $ is 
described {\em locally} by a hypersurface of degree $h$ 
\beqa
ord_{z(\Omega_t)}\, \theta (\cdot, \Omega_t)&=&
h^0(c_t, \widetilde{\cL}|_{c_t})
\eeqa

\appendix

\section{\label{spectral bundles}Moduli spaces of $SU(n)$ bundles}

\resetcounter

The moduli space of semistable 
$SU(n)$-bundles $V_F$ on an elliptic curve $F$ 
(so $V_F$ decomposes as a sum of flat line bundles $L_i=\cO_F(q_i-p_0)$
with $\sum q_i=0$ in the group law)
is a projective space,
encoding the $q_i$ as vanishing divisor of a meromorphic function $w$
\beqa
\M_F(V_F)\; & \cong & \; {\bf P}H^0(F, \cO_F(np_0))\; \cong \; {\bf P^{n-1}}
\eeqa
from (in affine coordinates of $z=1$; 
this is for $n$ odd, otherwise the last power is $x^{n/2}$)
\beqa
\label{locus}
w=a_0+a_2x+a_3y + \dots + a_n x^{\frac{n-3}{2}}y=0
\eeqa

Now one proceeds [\ref{FMW}] to the Calabi-Yau space $X$,
elliptically fibered over the base surface $B$ with fibre $F$
and section $\si$.
Using the line bundle $\T=K_B^{-1}$ over $B$ one has 
the Weierstrass representation\footnote{where 
$x,y,z\in H^0(X, \cO_X(3\sigma)\ox \T^i)$ with $i=2,3, 0$ 
and $g_2, g_3\in H^0(X, \T^i)$ with  $i=4,6$} 
of $X$ over $B$ (with the zero point $p=(0,1,0)$)
\beqa
\label{E fibration}
zy^2 & = & 4x^3 - g_2 xz^2 - g_3z^3
\eeqa 
This represents
$X$ as a hypersurface in a fourfold $W$ which is ${\bf P^2}$-fibered over $B$
\beqa
W\; & = & \; {\bf P}(\T^2\oplus \T^3\oplus \cO)
\eeqa
(with $x, y, z$ as homogeneous coordinates).
A polynomial representation of $\E=\pi^{-1}(b)$ arises just by restriction
of the Weierstrass fibration of $X$ over $B$ to that of $\E$ over $b$.

An $SU(n)$-bundle over $X$
(which is fibrewise suitably generic) determines a section of a 
${\bf P^{n-1}}$-bundle $\cW$ over $B$
(a relative projective space over $B$) and
has the $a_i\in H^0(B,\T^{-i})$ as homogeneous coordinates 
(more precisely $a_i\in H^0(B, \cO_{\cW}(1)\ox \T^{-i})$) on the total space
\beqa
\cW\; & = & \; {\bf P}\Big(\cO_B \oplus \bigoplus_{k=2}^n \T^{-k}\Big)
\eeqa
This gives for the moduli space of fibrewise semistable bundles a map
\beqa
\label{moduli space fibration map}
\M_X(V_X)\; & \lra & \; \Gamma(B, \cW)\; = \; 
{\bf P}H^0\Big(B, \cO_B \oplus \bigoplus_{k=2}^n \T^{-k}\Big)
\eeqa
For $s\in \Gamma(B, \cW)$
a further globalization datum is given by the twist line bundle 
$\M:= s^*(\cO_{\cW}(1))$ over $B$ (of $c_1(\M)=\eta$, say). 
Under $s$ the $a_i$ pullback to sections in $a_i\in H^0(B, \M \ox \T^{-i})$.
The $a_i$ give via $[(a_0, \dots , a_n)]$ 
the section in (\ref{moduli space fibration map}), cf.~[\ref{FMW}].

\subsection{\label{spectral equation}
The spectral surface and the spectral curve}

The spectral surface $C$ is given by (\ref{locus})
with $w\in H^0(X, {\cal O}(\sigma)^n \ox \pi^*\M)$ where
\beqa
\label{explicit coordinate forms}
\un{n=2}& \hspace{1.5cm}& w\, = \, a_0z+a_2 x\nonumber\\
\un{n=3}& \hspace{1.5cm}& w\, = \, a_0z+a_2 x+a_3 y
\;\;\;\;\;\;\;\;\;\;\;\;\;\; 
(\, = \, C_r z + B_{r-2}x + A_{r-3}y\, )\nonumber\\
\un{n=4}& \hspace{1.5cm}& w\, = \, a_0z^2 + a_2 xz + a_3 yz + a_4 x^2 
\nonumber\\
\un{n=5}& \hspace{1.5cm}& w\, = \, a_0z^2 + a_2 xz + a_3 yz + a_4 x^2 + a_5 xy
\nonumber\\
\un{n=6}& \hspace{1.5cm}& w\, = \, a_0z^3 + a_2 xz^2 + a_3 yz^2 + a_4 x^2z 
+ a_5 xyz + a_6 x^3
\eeqa
or in general (where $0\leq i, j$ and $j\leq 1$; $w$ has degree 
$[\frac{n}{2}]=
\left\{ \begin{array}{ll}
\frac{n}{2}\;\;\;\;\; n \, \mbox{even}\\
\frac{n-1}{2}\;\; n \, \mbox{odd}
\end{array} \right. $ in $x,y,z$)
\beqa
w & = & \sum_{\stackrel{\scriptstyle m=0}{2i+3j=m}}^n
a_m x^i y^j z^{[\frac{n}{2}]-(i+j)}\\
&=&a_0 z^{[\frac{n}{2}]} + a_2 x z^{[\frac{n}{2}]-1} 
+ a_3 yz^{[\frac{n}{2}]-1} + \dots \nonumber\\
&&
\;\;+\;\left\{ \begin{array}{ll}
a_{n-2}x^{[\frac{n}{2}]-1}z + a_{n-1}x^{[\frac{n}{2}]-2}yz \, + \;
a_n \, x^{[\frac{n}{2}]}\;\;\;\;\;\;\;\;\;\;\;\;\;\;\;\;\;\;\;\;\;\;\;
n \, \mbox{even}\\
a_{n-3}x^{[\frac{n}{2}]-1}z   + 
a_{n-2}x^{[\frac{n}{2}]-2}yz \, + a_{n-1}x^{[\frac{n}{2}]}
+a_n x^{[\frac{n}{2}]-1}y \;\;\; n \, \mbox{odd}
\end{array} \right. 
\nonumber
\eeqa

For the elliptic curve $F\subset {\bf P^2}$ considered in (\ref{E fibration}), 
the divisor $(z)=l\subset {\bf P^2}$ becomes $(z)|_F=3p_0$ on $F$ where
$p_0=(0,1,0)$. To encode $n$ points on $F$ (for $n$ even, say) one 
chooses a homogeneous polynomial $w^{(hom)}_{n/2} (x,y,z)$ of degree $n/2$. 
From its $3 n/2$ zeroes on $F$ only $n$, say $q_i$,
carry information as $n/2$ of them are always at $p_0$: the rewriting
$w^{(hom)}_{n/2}(x,y,z)=z^{n/2}w^{aff}_{n/2}(x/z,y/z)$ shows 
$3 n/2$ zeroes at $p_0$, $n$ poles at $p_0$ and $n$ zeroes
at the $q_i$ ($x/z$ and $y/z$ have a pole at $p_0$ of order
$-(1-3)=2$ and $-(0-3)=3$, resp.).

This gives on $F$ for the divisor of $w^{(hom)}=w$
resp. for the divisor of zeroes of $w^{aff}$
\beqa
(w|_F)=\frac{n}{2}\sigma + \sum q_i\;\;\;\; , \;\;\;\;
\;\;\;\; (w^{aff}|_F)_0=\sum q_i
\eeqa

Restricting from the elliptic threefold $X\subset W$ 
(where the fourfold $W$ is the ${\bf P^2}$ bundle of Weierstrass 
coordinates over $B$) to the elliptic surface $\E$ one gets again 
(\ref{E fibration}) where now $g_2$ and $g_3$ are homogeneous 
polynomials of degree $4$ and $6$, respectively, on $b={\bf P^1}$.
Similarly one gets equation (\ref{locus}) for $c\subset \E \subset \cW_{b}$
(inside the threefold $\cW_{b}$ given by the ${\bf P^2}$ bundle 
over $b$) where $\T|_b=\cO_b(\chi)$
with $x, y, z\in H^0(\E, \cO_{\E}(3s+i\chi F))$ for $i=2, 3, 0$ and
$a_i|_b\in H^0(b, \cO_b(r-i\chi))$ where $\M|_b=\cO_b(r)$ and $r:=\eta\cdot b$.
The coefficients of the homogeneous polynomials $a_i$ can 
be considered as moduli of external motions of $c$ in $\E$. 
We set $\chi=1$ as this is the case of $b$ isolated (cf.~sect.~3.1 of
[\ref{extrinsic}]).

\subsection{\label{dimension}The dimension of the moduli space}

The moduli decomposition in (\ref{fibration structure})
can also be seen from the action of the canonical involution $\tau$ 
(inversion on the fibers,trivially on the base; 
it sends $y$ to $-y$ in Weierstrass coordinates). 
The base and (continuous) fibre moduli constitute the even and odd parts,
respectively, cf.~[\ref{CD}].
Considering the number $n_e/n_o$ of $\tau$-even/odd bundle moduli one has then
$n_e-n_o=h^{2,0}(C)-h^{1,0}(C)$, cf.~[\ref{CD}]. Considering [\ref{FMW}]
the $\tau$-equivariant index $\chi_{\tau}(Z, ad\, V)=
\sum_{i=0}^3(-1)^i\, \mbox{Tr}_{H^i(X, ad\, V)}\tau$
(as the ordinary index $\chi(Z, ad\, V)$ vanishes by Serre duality)
one gets (using the projector $(1+\tau)/2$ such that 
$I=-\sum_i (-1)^i h^i_{even}$)
\beqa
I&=&-\frac{1}{2}\chi_{\tau}(Z, ad\, V)=
-\sum_{i=0}^3(-1)^i\, \mbox{Tr}_{H^i(X, ad\, V)}\frac{1+\tau}{2}=n_e-n_o
\nonumber
\eeqa
(for trivial unbroken gauge group, say, such that $h^0_{even}$ and
$h^3_{even}=h^0_{odd}$ are zero; note that
Serre duality interchanges the even and odd subspaces of the respective
cohomologies, the canonical bundle being odd).
This gives [\ref{FMW}]
\beqa
I&=& rk - \sum_{i=1,2} \int_{U_i}c_2(V)\; =\; n-1 - 4\omega_{\gamma=0}+\eta c_1
\eeqa
where the $U_i$ are the two fixed point sets $\si$ and $\{y=0\}$.
Alternatively one gets [\ref{CD}]
\beqa
I+1\; =\; \chi\big(X, \cO(C)\big)&=&\frac{1}{12}\Big( c_2(C)+c_1^2(C)\Big) C 
\; = \;
\frac{1}{12}\Big( c_2(X)C+2C^3\Big)\nonumber\\
&=&n+\frac{n^3-n}{6}c_1^2+\frac{n}{2}\eta(\eta-nc_1)+\eta c_1
\eeqa
(with $c_1:=c_1(B)$).
In our case of $h^{1,0}(C)=0$ one gets a formula
for the number $h^{2,0}(C)$ of external deformations of $C$ in $X$.
Alternatively one can compute this\footnote{assuming that $\eta-nc_1$
is not only effective but even positive (cf.~also [\ref{BDO}])} 
from the degrees of freedom in the coefficient functions $a_i$ in the
equation for the spectral cover 
(taking into account the possibility of an 
overall scaling of the defining equation) (cf.~also (\ref{Leray}))
\beqa
\label{count of par's}
\mbox{$\sharp$ of parameters in the $a_i$} = 
\sum_{\stackrel{i=0}{i\neq 1}}^n \chi(B, \M\otimes \T^{-i})  -1
=
\sum_{\stackrel{i=0}{i\neq 1}}^n h^0(B, \cO_B(\eta - i c_1))  -1
\;\;\;\;\;\;\;\;
\eeqa

\subsection{\label{data}Spectral cover bundles}

As a further datum describing $V$ beyond the 
surface $C$, which encodes $V$ just fiberwise,
one has a line bundle $L$ over $C$ with  $V=p_*(p_C^*L\otimes {\cal P})$
where $p(=p_X)$ and $p_C$ are projections to the respective factors
in $X\x_B C$ and ${\cal P}$ the suitable Poincare line bundle [\ref{FMW}].
$L$ arises in the simplest case as a restriction $L=\underline{L}|_C$ to $C$
of a line bundle $\underline{L}$ on $X$.

The line bundle $L$ is, for $C$ ample, encoded just by a
half-integral number $\lambda$ stemming from a twist class $\gamma$. 
This occurs as $c_1(V)=\pi_*\big(c_1(L)+\frac{c_1(C)-c_1}{2}\big)=0$ implies
\beqa
\label{c_1(L)}
c_1(L) \; = \; -\frac{1}{2}\Big(c_1(C)-\pi_{C*}c_1\Big)+\gamma \;
= \; \frac{n\si + \eta + c_1}{2}|_C+\gamma
\eeqa
(we omit the obvious pullbacks for the integral (1,1)-classes 
$\eta$ and $c_1=c_1(B)$ on $B$)
where $\gamma$ denotes the only generally given class in 
$ker\, \pi_{C*}:H^{1,1}(C)\ra H^{1,1}(B)$ 
\beqa
\gamma \; =  \; \Gamma|_C \;\;\;\; \mbox{with} \;\;\;\;
\Gamma  \; =  \; \lambda \, \Big(n\sigma-(\eta-nc_1)\Big)
\eeqa
(here $\Gamma$ is an element of $H^{1,1}(X)$;
as in [\ref{extrinsic}] we will always assume $\la > 1/2$).

This gives
precise integrality conditions for $\la \in \frac{1}{2}{\bf Z}$:
if $n$ is odd, then one needs actually $\la\in \frac{1}{2}+{\bf Z}$;
if $n$ is even, then 
$\la \in \frac{1}{2}+{\bf Z}$ needs $c_1 \equiv 0 \; \mbox{mod}\, 2$
and $\la \in {\bf Z}$ needs $\eta \equiv c_1 \; \mbox{mod}\, 2$.
The nontriviality of this parameter is crucial to get chiral matter [\ref{C}].

Under our assumption $h^{1,0}(C)=0$ line bundles on $C$
are characterised by their Chern classes. Therefore
one can define line bundles $\underline{\G}$ and $\G$ 
on  $X$ and $C$, respectively, by
\beqa
c_1(\underline{\G}) \; = \; \Gamma \;\; ,  \;\; c_1(\G) \; = \; \gamma
\eeqa
(when we want to make the $\lambda$-dependence explicit
we denote these by $\underline{\G}_{\lambda}$ and $\G_{\lambda}$).
Further we will use the corresponding divisor classes 
(modulo linear equivalence) $\underline{G}$ and $G$ with
\beqa
\label{G definitions}
\underline{\G} \; = \; \cO_X\big( \underline{G}\big)\;\; , \;\;
\G \; = \; \cO_C\big( G \big)
\eeqa
i.e. one has  $\underline{G}=\lambda (n\si - \pi^*(M + n K_B))$
and, for example, $G|_c=\la(ns-(r-n\chi)F)|_c$.

Note that all these considerations of $\underline{\G}$ and $\G$ apply strictly
only formally as the corresponding Chern classes will, 
taken alone for themselves, be only half-integral in general; 
only the full combination in (\ref{c_1(L)}) will be integral 
and define a proper line bundle. Similar remarks apply
to decompositions written below.

Explicitely one finds for the various incarnations of the spectral line bundle
\beqa
\label{cal L def}
\underline{L} & = & \Big( \cO_X(\si)^n \ox \pi^* \M \ox 
\pi^* K_B^{-1}\Big)^{1/2}\ox 
\underline{\G}\; = \; 
\cO_X\Big( \frac{C+\pi^*\, K_B^{-1}}{2}+\underline{G}\Big)
\\
L & = & K_C^{1/2} \otimes\pi_C^*K_B^{-1/2}\otimes \G
\eeqa

{\em The spectral data along the elliptic surface $\E$}

We define also a restriction $l:=\underline{L}|_{\E}$ to $\E$.
So one has the inclusions of line bundles
\beqa
L \;\;\;\;\;\; \hra \;\;\; \underline{L} \;\;\;\;\;
&\;\;\;\;\;\;\;\;\;\;\;\;\;\;\;&\;\;\;\;\;
\; l|_c \;\;\;\;\;\;  \hra  \;\;\;\; l
\nonumber \\
\da \;\;\;\;\; \;\;\;\;\;\;\;\;  \;\;\; \da \;\;\;\;\;
& \;\; \mbox{and} \;\;& \;\;\;\;\;
\; \da \;\;\;\;\;\;\;\;  \;\;\; \;\;\; \;\;\; \da 
\\
C \;\;\;\;\;\; \hra  \;\;\; X\;\;\;\;\;
& \;\;\;\;\; &\;\;\;\;\;
\; c \;\;\;\;\;\;\;  \hra  \;\;\;\; \E 
\nonumber
\eeqa

The crucial fact is that one has for spectral bundles
\beqa
\label{bundle projection}
V|_{B}=\pi_{C*} L  \;\;\;\;\;\;\;\;\;\;\;\;\;\;\; 
\mbox{such that}\;\;\;\;\;\;\;\; V|_b=\pi_{c*}\; l|_c
\eeqa

Let us define the following expression related to the restriction to $\E$
\beqa
\label{r}
r & := & \eta \cdot b \in {\bf Z}
\eeqa
So $C=n\si + \eta$ gives $c=ns+rF$. 
One has by $\eta \geq n c_1$ that $r\geq n\, \chi$. For $C$ ample
one gets even $r>n$, cf.~equ.~(3.12) of [\ref{extrinsic}].
$Pfaff\not\equiv 0$ needs $r\geq \frac{n\la+1/2}{\la - 1/2}$ 
by (4.28) of [\ref{extrinsic}], so in the $SU(3)$ case
one has even $r\geq 5$ for $\la=3/2$ (while just $r\geq 4$ for $\la > 3/2$).

One has with $\Gamma|_{\E}=\lambda\big(ns-(r-n\chi)F\big)$ 
and $\T|_b=K_B^{-1}|_b=\cO_b(\chi)$
that\footnote{\label{K_c footnote}The equality to the second line
in (\ref{degree equalities}) follows from
$l^2|_c \ox \cF^{-2}=\cO_{\E}\Big( (ns+rF) + \chi F\Big)|_c
=\cO_{\E}\Big( (ns+rF-kF)+2F\Big)|_c= K_c \otimes \pi_c^*K_b^{-1}$
using $K_c=(c+K_{\E})c$ and $K_{\E}=-kF=(\chi -2)F$.} 
(with $l|_c=L|_c$)
\beqa
\label{l def}
l&=&\underline{L}|_{\E}\; = \; 
\cO_{\E}\Bigg( \frac{c+\pi_{\E}^{-1}\cO_b(\chi)}{2}+\underline{G}|_{\E}\Bigg)
\; = \;
\cO_{\E}\Bigg( \frac{ns+(r+\chi)F}{2}\Bigg)\ox \underline{\G}|_{\E} 
\;\;\;\;\;\;\;\;\; \nonumber\\
& \Lra & l(-F) \; = \; \Big( K_{\E}\ox \cO_{\E}(c)\Big)^{1/2}
\ox \cO_{\E}\Big( ns - (r-n\chi)F\Big)^{\la}\\
\label{degree equalities}
l|_c&=&\cO_c\Bigg( \frac{ns + (r+\chi)F}{2}\Big|_c \Bigg)\otimes \cF =
\Big(K_C^{1/2} \otimes\pi_C^*K_B^{-1/2}\Big)\Big|_c \otimes \cF=
K_c^{1/2}\otimes\pi_c^*K_b^{-1/2}\otimes \cF\nonumber\\
&\Lra& \cL\; =\; l(-F)|_c\; = \; K_c^{1/2}\otimes \cF\; = \; 
\cO_{\E}(\al s + \beta F)|_c
\eeqa 
where
\beqa
\label{alpha beta definitions}
\al   \; = \; n \Big(\la + \frac{1}{2}\Big)\, , \;\;\;\;\;\;\;\;\;\;
\beta \; = \; \Big(n\la + \frac{1}{2}\Big)\chi-\Big(\la - \frac{1}{2}\Big)r-1
\eeqa
Note that our standing assumption $\chi=1$ gives these precise integrality
conditions for $\la\in \frac{1}{2}{\bf Z}$: for $n$ odd one has $\la \in 
\frac{1}{2}+{\bf Z}$, for $n$ even one has $\la\in {\bf Z}$ and $r$ odd.

Here we have also introduced the line bundle on $c$ 
given by the restriction\footnote{the line bundle $\cF$ {\em on $c$}
will exist without the proviso concerning
half-integrality because $K_c^{1/2}$ will exist (although with a
$Pic^0_{(2)}(c)$-ambiguity) while $\G|_c$ and $G|_c$ still must be read
with this proviso}
\beqa
\cF & = & \G|_c\; = \; \cO_c\big( G|_c\big)
\eeqa
One notes that the line bundle $\cF$ is flat as
\beqa
\label{ineffectivity}
\Big(ns-(r-n\chi)F\Big)(ns+rF)=0& \Lra & \deg \, G|_c=0
\eeqa

The question whether $b$ contributes to $W$ depends only on $V|_b$.
The class $\gamma$ does not occur for
a spectral bundle $V_{\E}$ over 
$b$ where $\pi_{c\, *}:H^{1,1}(c)\lra H^{1,1}(b)$ 
is injective, cf.~[\ref{FMW}]. This is in accord with the fact that 
$c_2(V|_{\E})=r$ sees only the $\gamma$-free part of
$c_2(V)=\eta \si + \omega $ 
and not the $\gamma$-dependent pullback-class $\omega$
(there are some further consequences along these lines not pursued here).

\subsection{\label{aux locus}An auxiliary locus ($R\subset \ov{R}$)}

To elucidate some of our examples of $SU(3)$ bundles we give a 
technical discussion of an auxiliary locus:
one has a second explicit locus of codimension $n-2$ ($R_i:= Res(a_i, a_n)$)
\beqa
\label{R relation}
(\R\subset)\;\; R:=\bigcap_{i=2, \dots , n-1}\, (R_i)
\eeqa
For $r=n+1$ (where $\deg a_n=1$ and $j=1$ in (\ref{curly R relation})) 
one has $\R=\R^1=R$ (cf.~below).

In our consideration above we compared two loci: the structurally defined 
locus $\Sigma$ and its sublocus $\R$ which is defined by explicit equations.
In direct analogy to the discussion in (\ref{Sigma relation}) and 
(\ref{curly R relation}) let us define the following loci 
\beqa
( \Sigma \;\; \subset ) 
\;\;\;\; \;\; \Sigma^1&:=& \Big\{t\in \M_{\E}(c)\, \Big| \, 
(ns-F)|_{c_t}\sim \mbox{effective}\Big\}\;\;\;\\
\cup \;\;\;\;\;\;\;\;\;\;\;\;\;\; \cup&&\nonumber\\
\label{slash R and slash R1 relation}
( \R \;\; \subset ) \;\;\;\;\;\;  \R^1&:=& \Big\{t\in \M_{\E}(c)\, \Big| \, 
(ns-F)|_{c_t}= \mbox{effective}\Big\}\;\;\;
\eeqa
so $t\in \R^1$ just if $F_{u_j}|_c=\{np_0,u_j\}$ for some $j$.
Concerning the locus $\Sigma^1$ one has 
\beqa
\Sigma^1&=&\left\{ \begin{array}{ll}
\M_{\E}(c) & \;\;\;\;\;\;
\mbox{for} \;\;  r\geq n\frac{n+1}{2} \\
\Sigma & \;\;\;\;\;\; 
\mbox{for} \; \; r = n+1
\end{array} \right.
\eeqa
(if $r=n+1$ is attainable, cf.~after (\ref{r})).
The second relation is obvious; the first one follows from the R-R
computation $h^0(c_t, \cO_{\E}(ns-F)|_{c_t})=n(r-n-1)-n(r-\frac{n+1}{2})
+h^1(c_t, \cO_{\E}(ns-F)|_{c_t})=h^0(c_t, \cO_{\E}(rF)|_{c_t})-n\frac{n+1}{2}$
together with $h^0(c_t, \cO_{\E}(rF)|_{c_t})\geq r+1$ from the long exact
sequence associated with the restriction sequence from $\E$ to $c_t$.

\noindent
{\em \fbox{$SU(3)$ bundles}}

In view of the examples mentioned we study more closely the case $n=3$.
Here one gets (where we use already that that $\R=\Sigma$ for $n=3$,
cf.~(\ref{divisor equality}); for $r\geq 6$ use R-R)
\beqa
\Sigma^1&=&\left\{ \begin{array}{lll}
\M_{\E}(c) & \;\;\;\;\;\;
\mbox{for} \;\;  r\geq 6 \\
\R^1=R & \;\;\;\;\;\;
\mbox{for} \;\;  r=5 \\
\Sigma=\R=\R^1=R & \;\;\;\;\;\; 
\mbox{for} \; \; r = 4
\end{array} \right.
\eeqa
(where $r=4$ needs $\la > 3/2$, cf.~after (\ref{r})). Here for $r=5$ the 
non-triviality of $H^0(c_t, \cO_{\E}(3s-F)|_{c_t})$ is controlled by 
$\det \bar{\iota}_1$ where $\bar{\iota}_1$ is the map in the long exact 
sequence
\beqa
0\lra H^0\Big(c_t, \cO_{\E}(3s-F)|_{c_t}\Big) \lra 
H^1\Big(\E, \cO_{\E}(-(r+1)F)|_{c_t}\Big) \stackrel{\bar{\iota}_1}{\lra}
H^1\Big(\E, \cO_{\E}(3s-F)|_{c_t}\Big) \nonumber
\eeqa
Here the second term is the five-dimensional space
$H^1(b, \cO_b(-(r+1)))\cong H^0(b, \cO_b(r-1))^*$
and the third term is the five-dimensional space
$H^1(b, \cO_b(-1)\oplus \bigoplus_{i=2}^3 \cO_b(-1-i))
\cong H^0(b, \cO_b(1)\oplus \cO_b(2))^*$. Now, here $R=(R_2)$ and 
the concrete expression for
$\det \bar{\iota}_1$ from equ.~(G.13) in [\ref{extrinsic}] 
is just $R_2=Res(a_2, a_3)=Res(B_3, A_2)$ which controlles $\R^1$.

\section{\label{jacobian}Riemann theta function and Jacobian fibration}

\resetcounter

\subsection{\label{Prelim and theta}Preliminaries concerning abelian varieties}

Recall that a complex torus $M=V/\Lambda$ (with $V\cong {\bf C^g}$ 
and $\Lambda \cong {\bf Z^{2g}}$ a discrete lattice) 
is an abelian variety, i.e. an (projective) 
algebraic variety (allowing a projective embedding), exactly if it admits
a K\"ahler form $H$ in the rational cohomology (coming from the basis 
$dx_i$ dual to the integral $\Lambda$-basis of topological cycles, cf.~below),
a so-called Hodge form (by averaging over the compact space $M$
one can always consider just invariant forms of this type);
the choice of cohomology class of $H$, 
or the corresponding datum of an ample line bundle
$L$ on $M$, is then called a polarization.

There are two bases on $H^1(M, {\bf C})$: 
if $z_i, i=1, \dots, g$ are Euclidean coordinates one gets 
the corresponding $1$-forms $dz_i$ and $d\bar{z}_i$; on the other hand
if $\la_i, i=1, \dots, 2g$ are elements of an integral basis of $\Lambda$
they build also a basis for the real vector space $V$; this gives 
dual real coordinates $x_i, i=1, \dots, 2g$ on $V$ and corresponding 
$1$-forms $dx_i, i=1, \dots, 2g$ with $\int_{\la_j}dx_i=\delta_{ij}$
for $\la_j\in \Lambda\cong H_1(M, {\bf Z})$.
If $e_i (i=1, \dots, g)$ is a complex basis of $V$
the $g\x 2g$ period matrix $\Omega'$ effects via $dz_i=\Omega'_{ij}dx_j$ (and 
the conjugate for $d\bar{z}_i$) the change between the two mentioned bases.
In the topological basis one gets for the integral (invariant) $2$-form $H$
(where the $\delta_{\al}$ are integers with $\delta_{\al} | \delta_{\al +1}$)
\beqa
H & = & \sum \delta_{\al}\, dx_{\al}\wedge dx_{g+\al}
\eeqa
$H$ is called a principal polarization
if the elementary divisors $\delta_{\al}$ are all equal to $1$.
$(M,L)$ is then called a principally polarised abelian variety (p.p.a.v.).
Equivalently this means that the corresponding line bundle
$L$ has an essentially unique non-trivial section, i.e. $h^0(M, L)=1$.
Precisely then the period matrix can (using $e_i=\delta_i^{-1}\la_i$)
be normalized to 
\beqa
\Omega' & = & (I, \Omega)
\eeqa
where $\Omega$ is an element of the Siegel upper half plane $\cH_g$
(symmetric matrices of positive definite imaginary part). 
Conversely, given a point $\Omega \in \cH_g$,
one gets the complex torus $M_{\Omega}={\bf C}^g/\Lambda_{\Omega}$
from the lattice $\Lambda_{\Omega}$
generated by the columns of $(I, \Omega)$.

A line bundle $L$ on $M$ arises as quotient of $V\x {\bf C}$ under
$(z,w)\sim (z+\la, e_{\la}(z)w)$ with a system of non-vanishing holomorphic 
functions $(e_{\la})_{\la \in \Lambda}$ 
with $e_{\la_2}(z+\la_1)e_{\la_1}(z)=e_{\la_1+\la_2}(z)$.
For $z_i$ coordinates on $V$ dual to a basis $e_i$ 
and $L$ defined by $e_{\la_{\al}}=1, e_{\la_{g+\al}}=\exp (-2\pi i z_{\al})$
sections $\tilde{\theta}$ are given by functions $\theta$ 
changing by the corresponding multiplier under $\la_{\al}$-shifts in $z$. 
Then $c_1(L)=[H]$ determines $L$ up to translation and the translates 
$\tau^*_{\mu}L$ are just the line bundles with the same multipliers,
except that the $e_{\la_{g+\al}}$ are muliplied 
by constants $c_{\al}=\exp(-2\pi i \mu_{\al})$:
for $\mu=\frac{1}{2}\sum \Omega_{\al \al}e_{\al}$, say, one has the
transformation laws
\beqa
\theta(z+\la_{\alpha})=\theta(z)\, , \;\;\;\;\;\;
\theta(z+\la_{g+\alpha})=
\exp\Big(-2\pi i (\frac{1}{2}\Omega_{\alpha\alpha}+z_{\alpha})\Big)\theta(z)
\eeqa
This characterises $\theta$ (up to a factor): one has 
(we consider principal polarizations $H$)
$H^0(M, L)={\bf C}\theta $ where 
(here $\theta$ is considered as a function on ${\bf C^g}\x \cH_g$)
\beqa
\label{theta series expansion}
\theta(z, \Omega)&=&\sum_{l\in {\bf Z^g}} 
\exp\Big\{ \, 2\pi i \, \Big(\frac{1}{2}\langle l, \Omega l\rangle 
+\langle l,z\rangle\Big)\Big\} 
\eeqa
(with $\langle \cdot, \cdot \rangle$ the canonical scalar product 
on ${\bf C^g}$).
$\theta(\cdot, \Omega)$ 
is the Riemann theta-function (an even holomorphic function)
of divisor $\Theta=\Theta_{\Omega}$ 
(the quasi-periodicity of $\theta$, considered as a function, 
shows that its zeroes are well-defined on $M_{\Omega}$).
We denote a translate by $\Theta_{\la}:=\Theta + \la$.
If $\Theta$ is the zero-divisor $(\tilde{\theta})$ of a section
$\tilde{\theta}$ of $L$ one has $\Theta=c_1(L)$ such that
\beqa
L\; & = & \; \cO_M(\Theta)
\eeqa
Conversely, any principally polarised abelian variety $(M,L)$
is of the form $(M_{\Omega}, \Theta_{\Omega})$.
The moduli space\footnote{$(M,L)$ and $(M',L')$ 
are called isomorphic if an isomorphism $f: M\ra M'$ exists with $f^*L'=L$.}
$\A_g$ of p.p.a.v.'s is a quasi-projective variety of dimension $g(g+1)/2$
\beqa
\A_g & = & \cH_g/Sp(2g, {\bf Z})
\eeqa

\subsection{\label{Jacobi}The Jacobian 
and Jacobi's inversion of Abel's theorem}

For $c$ a curve $\Omega'$ gives the integrals of a basis
$(\omega_{\al})\in  H^0(c, \Omega^1)$ of holomorphic differentials
over a symplectic basis of topological cycles in $H_1(c, {\bf Z})$ 
and the Jacobian $Jac(c)=H^0(c, \Omega^1)^*/H_1(c, {\bf Z})$ 
is a prinipally polarized abelian variety (essentially by Poincare duality;
the polarization can be identified with the intersection form of $c$).

The moduli space of curves of genus $g$ is a quasi-projective variety 
$\M_g$ of dimension $3g-3$.
Considering $c$ together with a symplectic basis of $H_1(c, {\bf Z})$
gives a topological cover $\tilde{\M}_g$ as moduli space
and the period map (giving $\Omega$) defines a holomorphic map
\beqa
\tilde{\M}_g & \stackrel{\tilde{P}}{\lra} & \cH_g
\eeqa
%given by integration over the $b$-cycles. 
Let us denote its image, the jacobian locus, by $\J_g$.
$\tilde{P}$ clearly descends to a map
\beqa
\M_g & \stackrel{P}{\lra} & \A_g
\eeqa
Here the map $P$ is injective by Torelli's theorem, 
thus providing the identification $\M_g\cong \J_g/Sp(2g, {\bf Z})$.
The curve $c$ can thus be recovered from $(Jac(c), \Theta_c)$.

$\theta$ lives naturally on the 
universal Jacobian bundle\footnote{here, and when including 
spin structures below, we will have no need to include
possible subtleties, for example stemming from curves with automorphisms, 
explicitly in the discussion ($\underline{Jac}$ is used naively)}
$\underline{Jac}$, 
the total space of the fibration
\beqa
\label{Jacobian fibration}
\underline{Jac} \;\;\;\;\;\;\;\;\;\; \nonumber\\
\;\;\;\;\;\;\;\;\;\;\;\;\;\;\;
p\, \da \; Jac(\cdot)\\
\M_g\;\;\;\;\;\;\;\;\; \nonumber
\eeqa

Let $\mu: c \lra Jac(c)$ be given by the integrals of the $\om_i$
from a fixed point $p_0\in c$ to $p\in c$. One has $\mu(c)\cdot \Theta = g$,
the intersection consisting of $g$ points (counting multiplicities)
if not $\mu(c)\subset \Theta$. It was Jacobi's insight that this
realizes an inversion to Abel's Theorem.
The latter was the assertion that the map 
$D\ra \mu (D)=(\sum_\la \int_{q_{\la}}^{p_{\la}} \om_i)_{i=1, \dots , g}$
for divisors $D=\sum (p_{\la}-q_{\la})$ of degree zero becomes
injective when working modulo linear equivalence, i.e. mediates
an injection of the $Pic_0$ group to the Jacobian.
Jacobi's inversion stated (fixing a reference point $p_0\in c$)
that for $\la \in Jac(c)$ a (generically uniquely determined)
effective divisor $D=\sum_{i=1}^g p_i\in c^{(g)}:=Sym^{g}c$ 
exists with $\sum (p_i-p_0)$ being mapped to 
$\la - \kappa$,
explicitely $ \{ p_i \} = \mu(c) \cap \Theta_{\la}$ if not 
$\mu(c)\subset \Theta_{\la}$ (or equivalently if not $\la = \kappa + \mu(D_g)$ 
with $h^0(c, D_g)>1$ where $D\in c^{(g)}$); so one has $W_g:=\mu(Sym^g c)
=Jac(c)$.

Riemann's theorem asserts that $\mbox{mult}_z\; \Theta  =  h^0(c, \cO_c(D_z))$
where codim  $\Theta_{sing}= 3$ if $c$ is not hyperelliptic 
(for $c$ hyperelliptic it is $2$). More precisely 
the codimension one analytic subvariety $\Theta_{-\kappa} = W_{g-1}$  
can be described near $z=\mu(D)$ by an equation $\det \, f_{ij} =0$
where $f_{ij}$ is an $h\x h$ matrix of functions holomorphic at $z$
where $h=h^0(c, \cO_c(D))$. For this recall for effective divisors $D$ the map 
$\mu: \; c^{(d)} \; \ra \; Jac(c)$ with $d=\deg\, D$, cf.~(\ref{mu map}).
Here one identifies the following kernel and cokernel of the induced 
differential map $d\mu$ on tangent spaces at $D$ and $z=\mu(D)$, respectively
\beqa
\label{Kempf}
0\, \lra \, T_D |D|\; \lra \; T_D c^{(d)}\;
\buildrel d\mu \over \lra \; T_{z=\mu(D)}Jac(c)\; \lra \; 
H^0(c, K_c -D)^*\, \lra \, 0
\eeqa
Taking dimensions here gives Riemann-Roch: 
$\dim \, |D| - d+g- h^0(c, K_c-D)=0$. For a basis $(f_i)$ 
of $\Phi(D):=\{\mbox{meromorphic} \; f:c\ra {\bf C}\, |\, (f)+D\geq 0\}$ 
the $f_i$ span the vector space of translations of the affine space $|D|$: 
an element $D^{\prime}\in |D|$ is given as $D+(f^{\prime})$
for an $f^{\prime}=\sum a_i^{\prime}f_i$.
Correspondingly then, for a basis $(\om_j)$ of $H^0(c, K_c -D)$, 
the $f^{\prime}\, \om_j$ span $H^0(c, K_c-D^{\prime})$ (the dual of the
cokernel of $d\mu$ at $D^{\prime}$) i.e.~their common vanishing 
characterizes $im \, d\mu$ in $T_{z^{\prime}=\mu(D^{\prime})}Jac(c)$; 
in other words, the $h\x (g-1-d + h)$ matrix of linear forms $(f_i\, \om_j)$, 
acting on $T_{z=\mu(D)} Jac(c)$, has all $h\x h$ minors
vanishing on the image in the tangent cone 
(Kempf proved that these equations are also sufficient). 
What was considered here infinitesimally 
can also be described locally, i.e.~as local equations for $d\mu$
which have the $(f_i\, \om_j)$ as linear parts.

\subsection{\label{theta characteristics}Theta characteristics}

Consider the set of spin bundles of $c$, i.e.~the set 
of square roots of the canonical bundle\footnote{using for the 
additive divisor class and the associated multiplicative
line bundle $K_c$ the same symbol}
\beqa
\cS p(c)&=&
\Big\{D\in (Div(c)/\sim) \Big|2D\sim K_c\Big\}
\; \cong \; \Big\{\cO_c(D)\in Pic(c)\Big| \cO_c(D)^2\cong K_c\Big\}\;\;\;\;\;\;
\eeqa
This set of socalled theta characteristics (or different spin structures)
is a $Pic_0^2(c)$-torsor, i.e.~its elements are just rotated through 
by this group of square roots of the trivial line bundle 
(which themselves lie in the degree zero component $Pic_0(c)$ of $Pic(c)$
whereas the theta characteristics have degree $g-1$); 
the set of these $2$-torsion points is isomorphic to
$({\bf Z}/2{\bf Z})^{2g}$ and has $2^{2g}$ elements;
the notation $K_c^{1/2}$ has this inherent $2^{2g_c}$-fold ambiguity.

Recall Riemann's theorem $W_{g-1}-\Delta=\Theta$ where 
$\Delta=\mu(D_0)-\mu((g-1)P_0)$. Sending
a theta characteristic $D$ to $W_{g-1}-\mu(D)$ establishes
an isomorphism to the set of those translates $\Theta_e$ which are
symmetric (which just comes down to $e\in Pic_0^2(c)$; here $D_0$ corresponds
just to $\Theta$ itself). This sends 
\footnotesize{$\Big(\!\begin{array}{c}a\\b\end{array}\!\Big)=
\Big(\!\begin{array}{c}\frac{1}{2}a'\\\frac{1}{2}b'\end{array}\!\Big)$}
\normalsize{$\in (\frac{1}{2}{\bf Z})^{2g}/{\bf Z}^{2g}$} to the zero locus of 
a shifted theta function (where $D=D_0+d$ with $\mu(d)=\Omega a +b$)
\beqa
\theta\Big[\begin{array}{c}a\\b\end{array}\Big](z, \Omega)
&=&
\sum_{l\in {\bf Z^g}} 
\exp\Big\{ \, 2\pi i \, \Big(\frac{1}{2}\Big\langle l+a,
\Omega (l+a)\Big\rangle 
+\Big\langle l+a,z+b\Big\rangle \Big)\Big\} \nonumber\\
&=&\exp\Big\{ \, 2\pi i \, 
\Big(\frac{1}{2}\Big\langle a, \Omega a\Big\rangle
+\Big\langle a,z+b\Big\rangle \Big)\Big\}
\theta(z+\Omega a + b, \Omega )
\eeqa
(cf.~(\ref{theta series expansion})).
$\cS p(c)$ is divided into two sets $\cS p_{\pm}(c)$
of $\frac{2^g(2^g\pm 1)}{2}$ even/odd structures 
\beqa
\cS p(c)&=&\cS p_{+}(c)\stackrel{\cdot}{\cup}\cS p_{-}(c)
\eeqa
according to the parity of 
$mult_D \Theta$ or equivalently $h^0(c, \cO_c(D))\equiv a'^{\, t}\cdot b' (2)$.
This parity stays constant in any family.
We denote the $2^{2g}$-section of $2$-torsion points of 
$p:\underline{Jac}\lra \M_g$ in (\ref{Jacobian fibration}) by
\beqa
\label{Z definition}
Z&=&\Big\{ \, [z]\in Jac(c)\cong {\bf C^g}/\Lambda\; \Big|\; 2[z]=[0], \, 
\Omega_c\in \M_g \, \Big\}\, = \, Z_+ \stackrel{.}{\cup}Z_-
\eeqa 
where $[z=\frac{1}{2}\Omega a'+\frac{1}{2}b']\in Z_{\pm}$ according to 
the parity of $a'^{\, t}\cdot b'$; so, this
$2^{2g}$-section $Z$ decomposes in a $2^{g-1}(2^g+1)$-section $Z_+$
and a $2^{g-1}(2^g-1)$-section $Z_-$.

For example, a nonsingular odd (with $h^0(c, \cO(D))=1$) theta 
characteristic $D\in Pic_{g-1}(c)$ (which always exists) corresponds to
\footnotesize{$\Big(\!\begin{array}{c}a\\b\end{array}\!\Big)$}
\normalsize{$\in (\frac{1}{2}{\bf Z})^{2g}/{\bf Z}^{2g}\cong Pic^0_2(c)$} 
where the shifted theta function has just a first order zero in $z$.
On a generic curve a theta characteristic has actually $h^0(c, \cO_c(D))=0$ 
or $1$. The locus of curves in $\M_g$ (for $g\geq 3$)
having an even theta characteristic with\footnote{or, equivalently, 
vanishing "theta-null", 
i.e.~$\theta[\!\begin{array}{c}a\\b\end{array}\!]
([0], \Omega_c)=0$ or $\theta(\Omega_c \, a + b, \Omega_c)=0$;
here the notion comes from "theta-nullwerte",
that is theta zero-values (i.e.~values at zero-argument in $z$)} 
non-vanishing $h^0(c, \cO_c(D))$  
is an irreducible divisor which we denote by $\M_g^1$ 
\beqa
\label{Z relation}
\M_g^1&=& \Big\{c\in \M_g\, \Big|\, \exists D\in \cS p_{+}(c) \;\; s.t. \;\; 
h^0(c, \cO_c(D))\neq 0\Big\}
\eeqa

So one has the following decomposition of $Z\cap \Theta$ and projection
of $Z_+\cap \Theta$
\beqa
\label{Z Theta decomposition}
Z\cap \Theta &=&(Z_+\cap \Theta) \; \stackrel{\cdot}{\cup} \; Z_- \nonumber\\
             & & \;\;\;\;\;\;\; \da \, p \\
             & & \;\;\;\;\; \M_g^1   \nonumber
\eeqa

\subsection{\label{Mathematical Standard Notation}Some Standard Notation}

We collect some mathematical standard notation and point to the place where
the corresponding notion is considered in greater detail.

$\theta(\cdot, \cdot)$ denotes the theta function, 
cf.~app.~\ref{Prelim and theta}.

$Jac(c)$ denotes the Jacobian of the curve $c$, cf.~app.~\ref{Jacobi}.

$Pic(c)$ denotes the group of line bundles over $c$; $Pic_m(c)$ denotes those
line bundles which are of degree $m$; $Pic_0^k(c)$ denotes the $k$-torsion
line bundles, i.e.~those $L$ for which $L^k\cong \cO_c$ holds.

$Sym^k c$ denotes the symmetric product of the point set $c$, i.e.~all 
$k$-tuples of points of $c$.

Div$(c)$ denotes the group of divisors on $c$; Div$_m(c)$ denotes those
divisors which are of degree $m$; Div$^{eff}(c)$ denotes the set of 
effective divisors, so in particular Div$_m^{eff}(c)=Sym^m(c)$.
Linear equivalence of divisors is indicated as usual by $\sim$,
so Div$_m(c)/\sim$ is isomorphic to $Pic_m(c)$.

$\mu: \mbox{Div}_m^{eff}(c)/\sim \; \lra Jac(c)$ denotes the Jacobi map, 
cf.~app.~\ref{Jacobi}, and $W_m$ its image; 
one has $W_{g-1}=\Theta + \mu(\frac{1}{2}K_c)$ according to Riemann's theorem
($g$ is the genus of $c$).

\section*{References}
\begin{enumerate}

\item
\label{FMW}
R. Friedman, J. Morgan and E. Witten, {\em Vector Bundles and F-Theory},
hep-th/9701162, Comm. Math. Phys. {\bf 187} (1997) 679.

\item
\label{FMWII}
R. Friedman, J. Morgan and E. Witten, 
{\em Principal G-bundles over elliptic curves},
alg-geom/9707004, Math.Res.Lett. {\bf 5} (1998) 97.

\item
\label{FMWIII}
R. Friedman, J.W. Morgan and E. Witten,
{\em Vector Bundles over Elliptic Fibrations}, alg-geom/9709029,
Jour. Alg. Geom. {\bf 8} (1999) 279.

\item
\label{Wittenhet}
E. Witten, {\em World-Sheet Corrections Via D-Instantons},
hep-th/9907041, JHEP {\bf 0002} (2000) 030.

\item
\label{W99}
E. Witten, 
{\em Heterotic String Conformal Field Theory And A-D-E Singularities},
arXiv:hep-th/9909229, 	JHEP {\bf 0002} (2000) 025.

\item
\label{Wfam}
C. Beasley and E. Witten,
{\em New Instanton Effects in String Theory},
arXiv:hep-th/0512039, JHEP {\bf 0602} (2006) 060.

\item
\label{C}
G. Curio, {\em Chiral matter and transitions in heterotic string models},
hep-th/9803224, Phys.Lett. {\bf B435} (1998) 39. 

\item
\label{CD}
G. Curio and R. Y. Donagi, {\em Moduli in N=1 Heterotic/F-Theory 
Duality}, hep-th/9801057, Nucl.Phys. {\bf B518} (1998) 603.

\item
\label{extrinsic}
G. Curio, {\em World-sheet Instanton Superpotentials 
in Heterotic String theory and their Moduli Dependence},
arXiv:0810.3087 [hep-th].

\item
\label{BDO}
E.I.~Buchbinder. R.~Donagi and B.A.~Ovrut, {\em Vector Bundle Moduli
Superpotentials in Heterotic Superstrings and $M$-Theory}, 
hep-th/0206203, JHEP {\bf 0207} (2002) 066.

\end{enumerate}
\end{document}